\documentclass{article}

\usepackage{PRIMEarxiv}

\usepackage[utf8]{inputenc} % allow utf-8 input
\usepackage[T1]{fontenc}    % use 8-bit T1 fonts
\usepackage{hyperref}       % hyperlinks
\usepackage{url}            % simple URL typesetting
\usepackage{booktabs}       % professional-quality tables
\usepackage{amsfonts}       % blackboard math symbols
\usepackage{nicefrac}       % compact symbols for 1/2, etc.
\usepackage{microtype}      % microtypography
\usepackage{lipsum}
\usepackage{fancyhdr}       % header
\usepackage{graphicx}       % graphics
\graphicspath{{media/}}     % organize your images and other figures under media/ folder

\usepackage{amsmath,amssymb}
\usepackage{bbm}
\usepackage{subfigure}
\usepackage{booktabs}
\usepackage{multirow}

\newcommand{\etal}{\emph{et al.} }
\newcommand{\ie}{{\emph{i.e.}}, }
\newcommand{\eg}{{\emph{e.g.}}, }

\newcommand{\gene}{\mathcal{G}_{enc}}
\newcommand{\gend}{\mathcal{G}_{dec}}
\newcommand{\dis}{\mathcal{D}}
\newcommand{\gen}{\mathcal{G}}

\title{Aligned Disentangling GAN: A Leap towards End-to-end Unsupervised Nuclei Segmentation}
\author{
  Kai Yao \\
  University of Liverpool\\
  School of Advanced Technology, Xi'an Jiaotong-Liverpool University. \\
   \And
  Kaizhu Huang \\
  Duke Kunshan University \\
  City\\
  \texttt{kaizhu.huang@dukekunshan.edu.cn} \\
   \And
   Jie Sun\\
   School of Advanced Technology\\
    Xi'an Jiaotong-Liverpool University\\
    \And
    Curran Jude\\
    University of Liverpool\\
    the United Kingdom\\
}

\begin{document}
\maketitle

\begin{abstract}
We consider  unsupervised cell nuclei segmentation in this paper. Exploiting the recently-proposed unpaired image-to-image translation between  cell nuclei images and randomly synthetic masks, existing approaches, \eg CycleGAN, have achieved encouraging results. However,  these methods usually take a two-stage pipeline and fail to learn end-to-end appropriately in cell nuclei images. More seriously, they could lead to the lossy transformation problem,  \ie the content  inconsistency between the original images and the corresponding segmentation output. To address these limitations, we propose a novel end-to-end unsupervised framework called Aligned Disentangling Generative Adversarial Network (AD-GAN).
Distinctively, AD-GAN introduces representation disentanglement to separate content representation (the underlying spatial structure) from style representation (the rendering of the structure). With this  framework, spatial structure can be preserved explicitly, enabling a significant reduction of macro-level lossy transformation. We also propose a novel training algorithm able to align the disentangled content in the latent space to reduce micro-level lossy transformation. Evaluations on real-world 2D and  3D datasets show that AD-GAN substantially outperforms the other comparative methods and the professional software both quantitatively and qualitatively.
Specifically,  the proposed AD-GAN leads to  significant improvement over the current best unsupervised methods by an average 16.1\% relatively (w.r.t. the metric DICE) on four cell nuclei datasets.  As an unsupervised method, AD-GAN even performs competitive  with the best supervised models, taking a further leap towards end-to-end unsupervised nuclei segmentation. Codes are available at: \href{https://github.com/Kaiseem/AD-GAN}{https://github.com/Kaiseem/AD-GAN}.
\end{abstract}
\keywords{Image segmentation \and Unsupervised learning \and Biomedical computing}
%\linenumbers
% \begin{IEEEkeywords}
% Image segmentation, Unsupervised learning, Biomedical computing
% %Pathology, Object segmentation,  Artificial neural networks
% \end{IEEEkeywords}
\section{Introduction}
Fluorescence microscopy image analysis, particularly automatic cell nuclei image segmentation, is essential for quantifying cell models efficiently and accurately. Supervised nuclei segmentation methods have achieved impressive results on fluorescence microscopy images~\cite{Falk2019}. However, due to variations of settings, annotation noises, and/or insufficient labelled data, these methods are usually difficult to be used in practice. %For this reason, recent research focus has shifted to unsupervised nuclei segmentation.
For this reason, the unsupervised nuclei segmentation methods have recently drawn much attention.

\begin{figure}[!th]
\centering
\includegraphics[width=0.8\columnwidth]{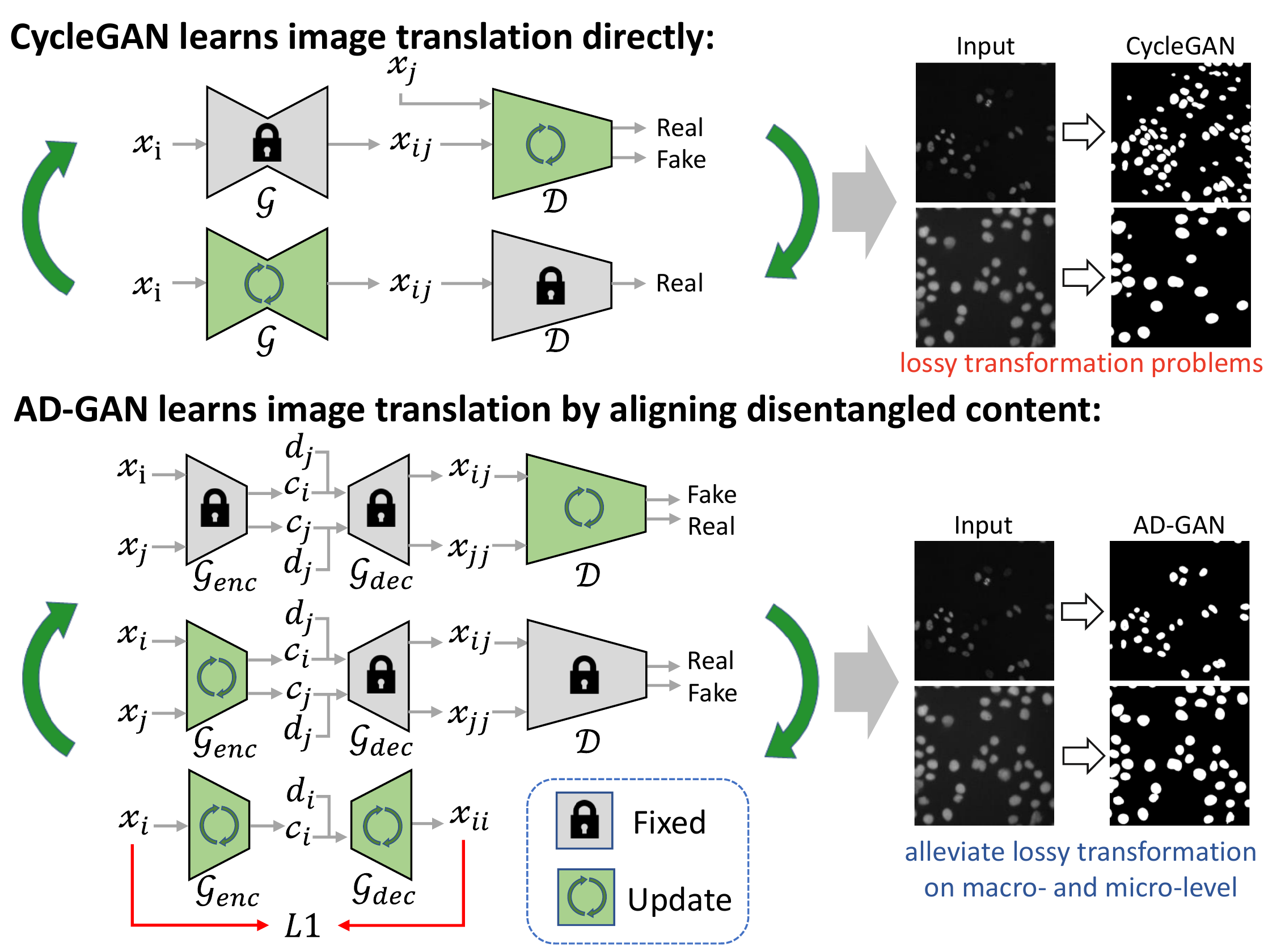}
\caption{Comparison of the conventional CycleGAN training scheme and the proposed AD-GAN training scheme during cross-domain translation. CycleGAN learns image translation directly by alternately freezing the generator and the discriminant, which may results in lossy transformation problem. In order to alleviate this, our proposed model engages a content-style disentangled auto-encoder structure and decouples the training of decoder from cross-domain translation. The encoder is forced to align the  the disentangled content during cross-domain translation so that the disentangled content from both domains can be decoded correctly by the same decoder.}
\label{fig:adt}
\end{figure}

The objective of unsupervised nuclei segmentation is to obtain the segmentation mask accurately via a one-to-one mapping from the input nuclei image without any annotation information. This problem can also be considered as unpaired image-to-image translation, meaning that no pair ground-truth is available between the nuclei image and the segmentation mask. Recent studies show that CycleGAN based unsupervised nuclei segmentation achieves the current state-of-the-art results. For instance, Böhland \etal investigated in a systematic study about the influence of synthetic masks' object properties on CycleGAN supported segmentation pipelines~\cite{Bohland2019Influ}. They first trained a CycleGAN to create paired synthetic data from the synthetic masks, and then learned a semantic segmentation model. Fu \etal  developed a spatial constrained CycleGAN (SpCycleGAN)~\cite{Fu2018three} to deal with the spatial offset problem in 3D images and then used synthetic paired dataset on supported 3D segmentation pipelines~\cite{DeepSynth}.

\begin{figure}[!th]
\centering
\includegraphics[width=0.85\columnwidth]{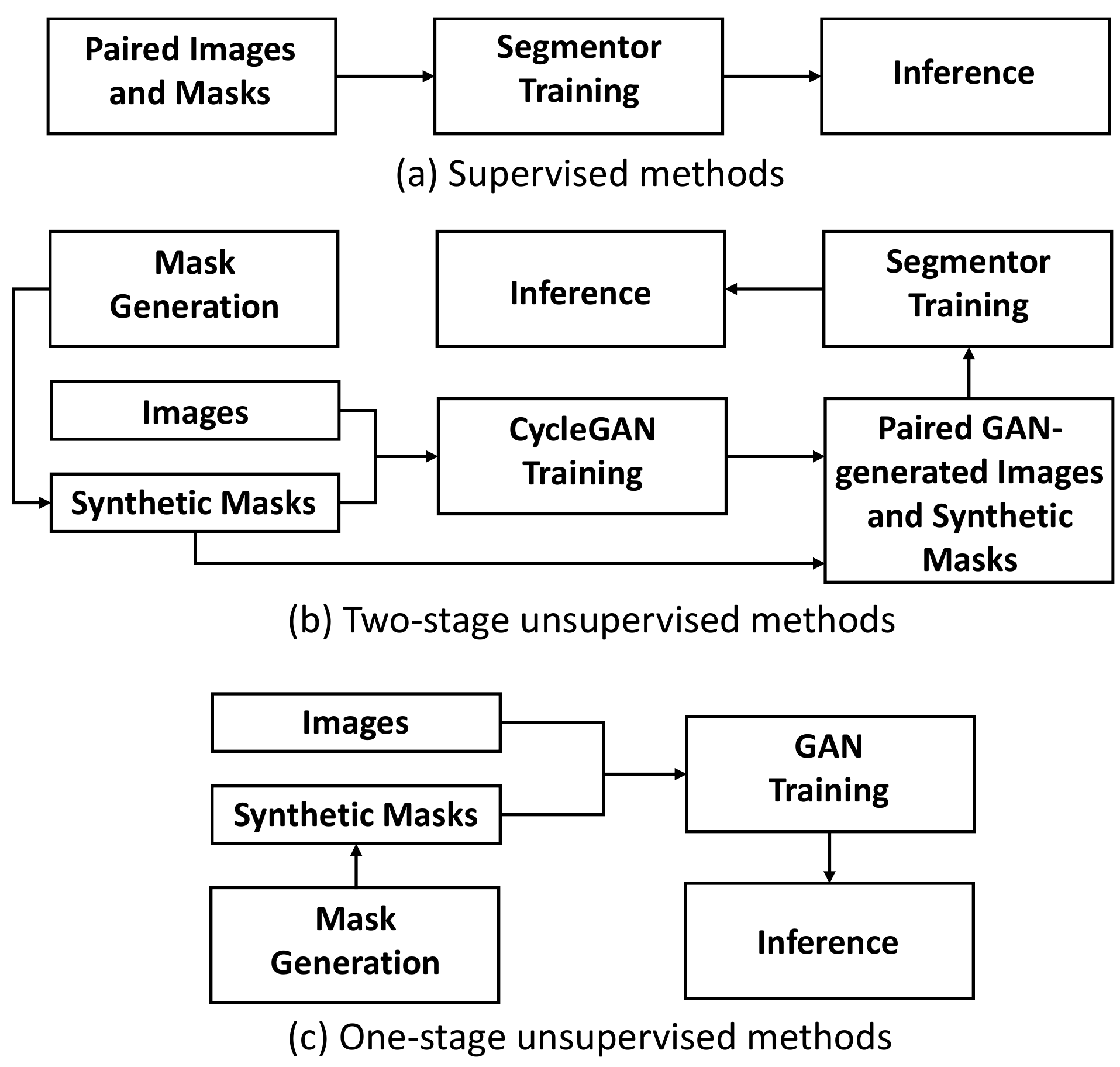}
\caption{Illustration of difference between supervised methods, two-stage CycleGAN-based pipeline of unsupervised methods, and one-stage GAN-based unsupervised methods.}
\label{fig:diff}
\end{figure}

Despite their success, these CycleGAN based methods are usually not trained end-to-end when used in nuclei cell segmentation, since the masks generated by the corresponding generator may contain too many errors. Instead, these methods often engage a two-stage pipeline trying to train a robust segmentor with the potentially erroneous CycleGAN-synthesized data. This would introduce extra time and computation cost, and significantly increase the system complexity, as shown in \autoref{fig:diff}.
% since a two stage pipeline shall
% they need generate firstly synthetic data so as to form the paired data.
More seriously, the so-called lossy transformation problem~\cite{Steganography} usually exists, \ie the content inconsistency between the original images and the corresponding segmentation (mask) output. Such inconsistency includes nuclei deletion/addition at the macro-level, and location offset, shape difference at the micro-level, which can been seen in the right-upper part of \autoref{fig:adt}. Though some recent methods~\cite{Fu2018three,Ihle2019UDCT,matchobj} have been proposed attempting to alleviate it by adding regularization terms or matching the exact global properties, their performance is still limited. Particularly, these recent frameworks all engage entangled representation models~\cite{CycleGAN}, making them difficult to separate the content (semantic parts to be segmented) from the style (semantic parts to be masked or removed).

In this work, we propose a novel end-to-end framework, called Aligned Disentangling Generative Adversarial Network (AD-GAN) to address the lossy transformation problem typically in unsupervised nuclei segmentation. Inspired by recent achievements on image-to-image translation~\cite{MUNIT,DRIT,DRIT_plus}, we take advantages of representation disentanglement and exploit the auto-encoder as the generator to extract disentangled representations. Different from the existing general-purpose representation disentangling models which learn a distinctive style for each sample, we assume that a single domain style can be learned for either the nuclei image domain or the synthetic mask domain, which holds for most nuclei images and synthetic segmentation masks since they all contain very simple textures. We also design a unified framework able to take one-hot domain labels to control translation directions. With this novel design, the disentangled content representation of the two domains can share the same latent space naturally. We further propose an aligned disentangling training strategy, capable of aligning the disentangled content representation in the latent space, as shown in the bottom part of \autoref{fig:adt}. The aligned content representation ensures one-to-one mapping not only at the image level but also at the semantic object level. Importantly, evaluations on 2D and 3D datasets show our method can substantially alleviate the lossy transformation problem and achieve significant improvements over the current best CycleGAN based methods.

Overall, the contributions of this paper are four-fold:
\begin{enumerate}
\item  We propose a novel end-to-end unsupervised nuclei segmentation framework called AD-GAN. We take a further leap and  achieve so far the best results with an improvement of  16.1\% averagely on both  2D and 3D data (even comparable with those of supervised methods);
\item We alleviate the macro-level lossy transformation problem in unsupervised nuclei segmentation by learning disentangled representations with a simpler yet effective architecture;
\item We propose a novel training strategy called Aligned Disentangling Training which can further align the disentangled content representation to reduce micro-level lossy transformation problem;
\item By better coping with the lossy transformation problem, our proposed method can be readily extended to instance segmentation and image synthesis.
% \item A new 3D fluorescence image dataset of 20 unlabelled images and one fully annotated image with more than 800 exhaustively annotated nuclei.
\end{enumerate}

\begin{figure*}[t]
\centering
\subfigure[CycleGAN]{
\includegraphics[width=0.20\columnwidth]{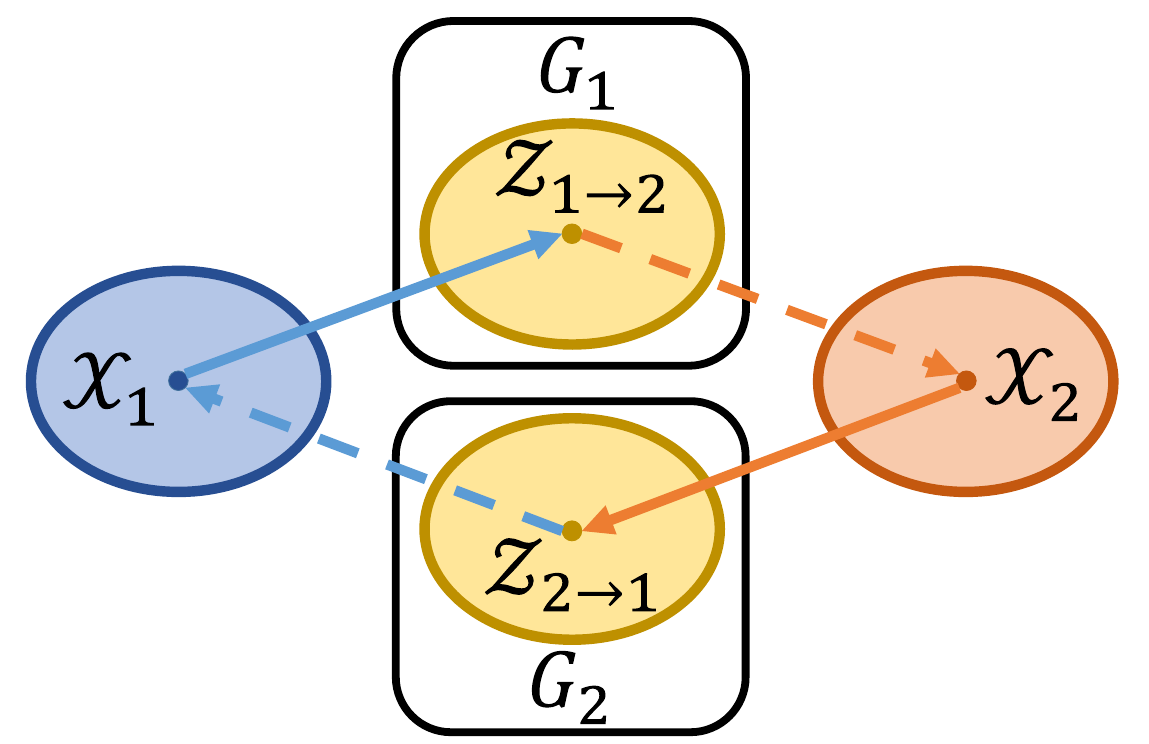}
\label{fig:2a}
}
\subfigure[MUNIT, DRIT, DRIT++]{
\includegraphics[width=0.20\columnwidth]{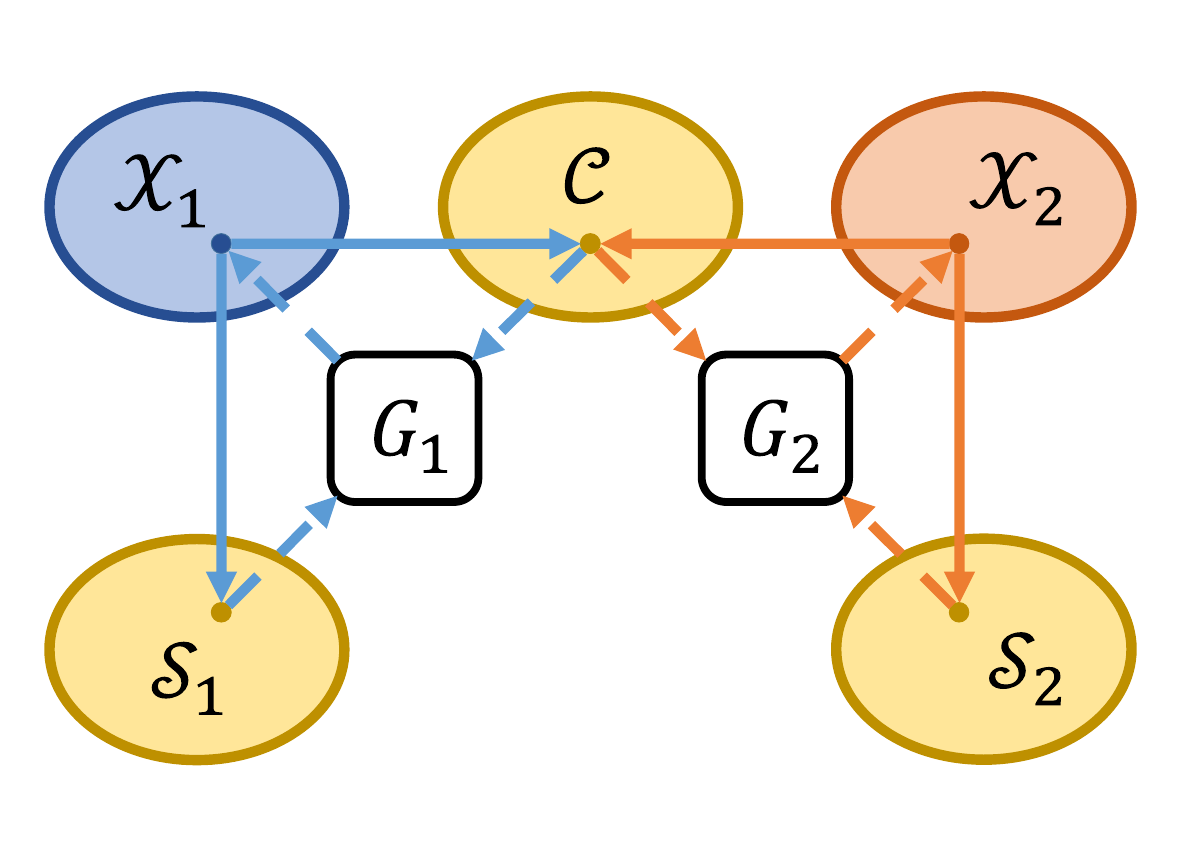}
\label{fig:2b}
}
\subfigure[AD-GAN]{
\includegraphics[width=0.20\columnwidth]{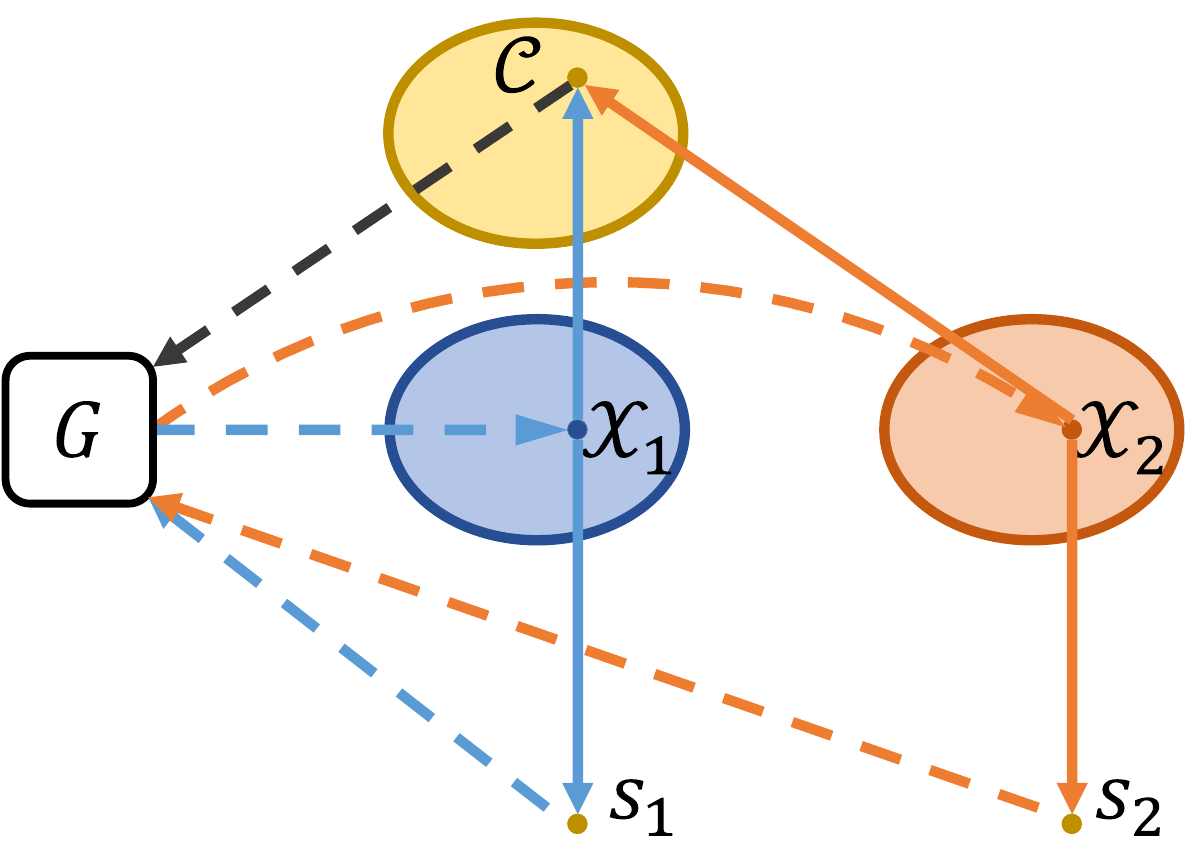}
\label{fig:2c}
}
\caption{Comparisons of unsupervised image-to-image translation models. Denote ${x}_{1}$ and ${x}_{2}$ as images in domain $\mathcal{X}_{1}$ and $\mathcal{X}_{2}$: (a)~CycleGAN maps ${x}_{1}$ and  ${x}_{2}$ onto separated latent spaces. (b)~Common representation disentangling models disentangle the latent spaces of ${x}_{1}$ and  ${x}_{2}$ into a shared content space $\mathcal{C}$ and a separate style space $\mathcal{S}$ for each domain. (c)~In our AD-GAN, one single style is assumed for each domain (which holds in nuclei cell image segmentation). AD-GAN can disentangle the latent spaces of ${x}_{1}$ and ${x}_{2}$ into a shared content space $\mathcal{C}$ and two single domain style representation ${s}_{1}$ and ${s}_{2}$. AD-GAN enjoys a more simple, accurate, and end-to-end structure.}
\label{fig:comparison}
\end{figure*}

\section{Related Work}
\label{sec:related-word}
Existing unsupervised nuclei segmentation methods~\cite{Ho2017NS,brieu2019domain,Hou2019RHI,Fu2018three,Bohland2019Influ} usually adopt a two-stage pipeline, \ie 1) they first generate synthetic data so as to pair the input images and the synthetic masks, and 2) they learn the segmentation model with the synthetic paired data. Able to exploit the cycle consistency to ensure a good bidirectional mapping as depicted in \autoref{fig:2a}, CycleGAN and its variants are typically used in the first stage.  However, the lossy transformation problem commonly exists in these methods, leading that the synthesized data could not be well paired with the input images. As such, these methods may not be trained end-to-end appropriately, making it difficult to be applied in practice.  To alleviate this drawback, SpCycleGAN~\cite{Fu2018three} was proposed with a spatial regularization term for avoiding the spatial offset problem; UDCT~\cite{Ihle2019UDCT} was also developed by adding a novel histogram discriminator to make the style transfer more reliable. Though some encouraging results have been achieved, these methods do not learn disentangled representations.  Consequently they cannot well separate the contents from the styles and fail to handle the lossy transformation problem properly.

Aiming to decompose contents and style explicitly, recent studies focused on learning disentangled representations~\cite{MUNIT,AugCGAN,DRIT,DRIT_plus}. By assuming that the domain-invariant representation refers to content and the domain-specific representation refers to style, these methods have obtained substantial improvement in general-purpose unpaired image-to-image translation. Unfortunately, when applied in unsupervised nuclei segmentation, these methods may not achieve satisfactory performance. In particular, they assume that the style is diverse within domain and learn distinctive style for each sample, as illustrated in \autoref{fig:2b}. However, nuclei cell images  typically contain simple style in both image domain and the mask domain. In this scenario, given too flexible styles,  contents (or nuclei objects) may not necessarily be represented by domain-invariant features; instead, some contents may be described by certain domain-specific features. Therefore, some contents (nuclei) could be incorrectly learned as styles and are then removed in the mask domain, resulting in the content inconsistency during cross-domain translation.

Different from the above mentioned methods, we assume that one single style can be learned for each domain, as consistently observed from the nuclei images and synthetic masks. This enables an efficient and simple disentanglement framework. Separating content from style efficiently, our proposed framework can not only be trained end-to-end, but also achieve remarkably better performance than the previous methods including the current popular professional software for unsupervised nuclei segmentation. It is noticed that, UFDN~\cite{UFDN} has a similar assumption to ours that each domain has one domain style. They use VAE to disentangle content features but it is not guaranteed that the features are fully disentangled due to lacking cycle consistency~\cite{DALDR}. This results in much worse performance than our proposed AD-GAN method. More details can be later seen in Section~\ref{sec:experimental-results}.
%existing state-of-the-art content and style disentangled methods are all GAN-based methods.

\section{Proposed Method}
\label{sec:propsedmethod}

% In this section, we firstly describe the the cause of lossy transformation in unsupervised nuclei segmentation tasks. Afterward, we then discuss how we handle this with the proposed AD-GAN and aligned disentangling training(ADT) as well as the loss function for optimizing the whole GAN network.

We solve the unsupervised nuclei segmentation as an unsupervised unpaired image-to-image translation problem in this paper. Two different domain images are considered, \ie the microscopy cell image domain $\mathcal{X}_{1}$ and the synthetic segmentation mask domain $\mathcal{X}_{2}$, which can be easily generated by inserting 2D ellipse or 3D ellipsoid structure with random rotations and translations.

We propose a novel GAN-based model termed as AD-GAN as illustrated in \autoref{fig:2c}. Specifically,  we design a novel framework to disentangle the latent spaces of images and masks into a shared content space $\mathcal{C}$ and two single style representations, one for each domain. With the same-style assumption within each domain, a disentanglement of content from styles can be efficiently accomplished by conditioning domain labels. This also enables us to design a novel unified single generator in contrast to two generators as typically required in the traditional disentanglement models (see \autoref{fig:2b} and \autoref{fig:2c} ). With a novel training strategy called Aligned Disentangling Training, the proposed framework can not only be trained end-to-end, but also it can ensure that the disentangled content representation of the two domains are readily aligned and mapped to the same latent space.

In the following, we will first focus on introducing the novel unified generator $\gen$ of AD-GAN while exploiting a domain discriminator~\cite{StarGANv2} in our design which is a standard Markovian discriminator~\cite{pix2pix} with two output branches. We then discuss how we design the novel aligned disentangling training as well as the theoretical interpretation of our method. Finally, we describe the loss function for optimizing the whole GAN network.

\begin{figure*}
\centering
\includegraphics[width=0.95\columnwidth]{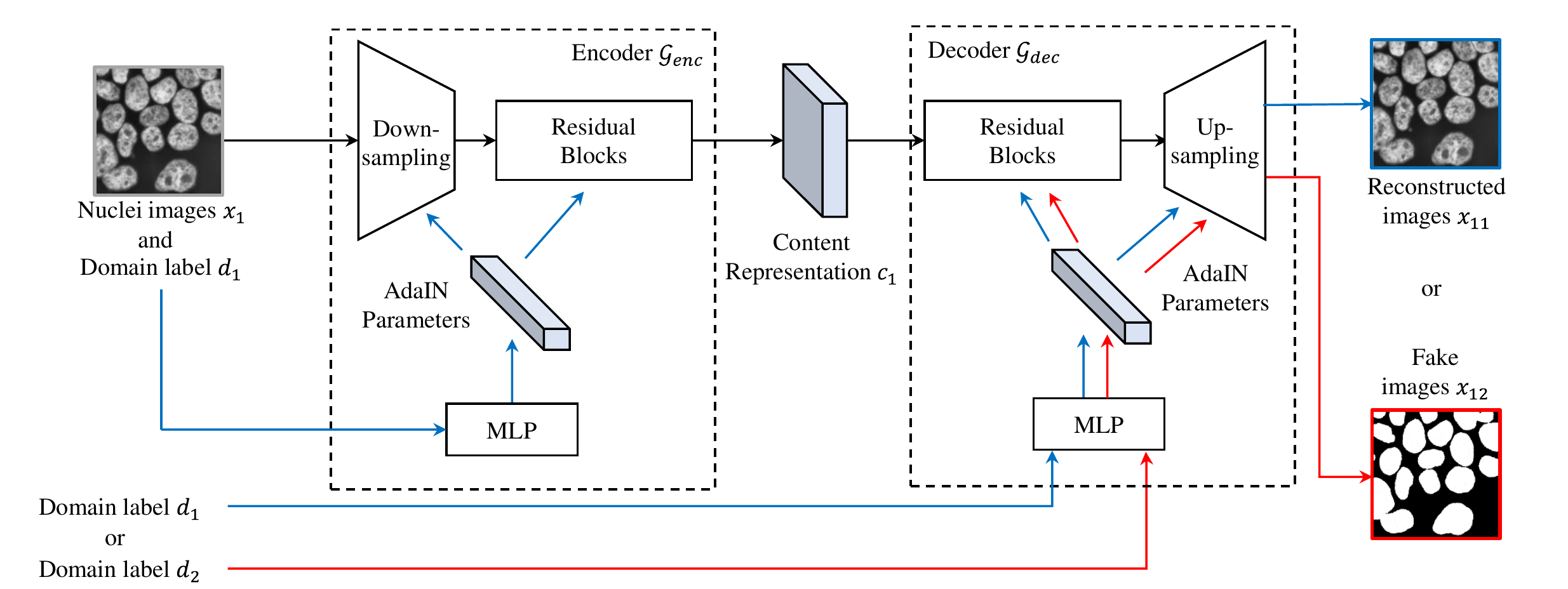}
\caption{Architecture of the proposed auto-encoder based generator. Under our within-domain same-style assumption in this task, the major style of each domain can be learned as style representation so that the disentangled content can be preserved explicitly.}
\label{fig:architecture}
\end{figure*}

\subsection{Architecture of Generator}
Let $x_i \in \mathcal{X}_{i}$ $(i=1,2)$ represent a sample from the cell image domain or synthetic mask domain, and $d_i$ denote the corresponding domain label. \autoref{fig:architecture} shows the architecture of our auto-encoder based generator $\gen$, which consists of an encoder $\gene$ and a decoder $\gend$. Different from MUNIT~\cite{MUNIT} which uses a style encoder to extract diverse style code, we use the domain label to replace the style code, so that one single style representation for each domain can be learned by a multi-layer perceptron (MLP). Inspired by recent work that uses affine transformation parameters in normalization layers to represent styles~\cite{MUNIT,StarGANv2}, we equip both the encoder and decoder with Adaptive Instance Normalization (AdaIN)~\cite{adain} layers to inject the style representations (see the appendix for details.). Moreover, we use a unified framework for both domains since the two learned style representations are discriminative. As shown in the blue path of \autoref{fig:architecture}  a content representation  $c_i$ can be factorized by the encoder conditioned on domain label $d_i$, which reflects the nature of nuclei images and masks in this task \ie $c_i =\gene(x_i,d_i)$. Simultaneously, the content representation  $c_i$ and the style representation learned from domain label $d_i$ can be composed into reconstructed image $x_{ii}$ by the decoder, \ie $x_{ii}=\gend(\gene(x_i,d_i),d_i)$. Image-to-image translation is performed by swapping decoder's domain label to another, as illustrated in the  red path of \autoref{fig:architecture}. Consequently, fake images $x_{ij}$ can be obtained, \ie $x_{ij}=\gend(\gene(x_{i},d_{i}),d_{j})$, $i \ne j$.

In addition, we adapt a domain discriminator~\cite{StarGANv2} in our design which is a standard Markovian discriminator~\cite{pix2pix} with two output branches. Each branch $\dis_i$ learns a binary classification determining whether an image $x_ii$ is a reconstructed image or a fake image generated by $\gen$, \ie $\dis_i(x_i)=\dis(x_i|d_i)$.

\subsection{Aligned Disentangling Training}
\begin{figure*}
\centering
\includegraphics[width=0.95\columnwidth]{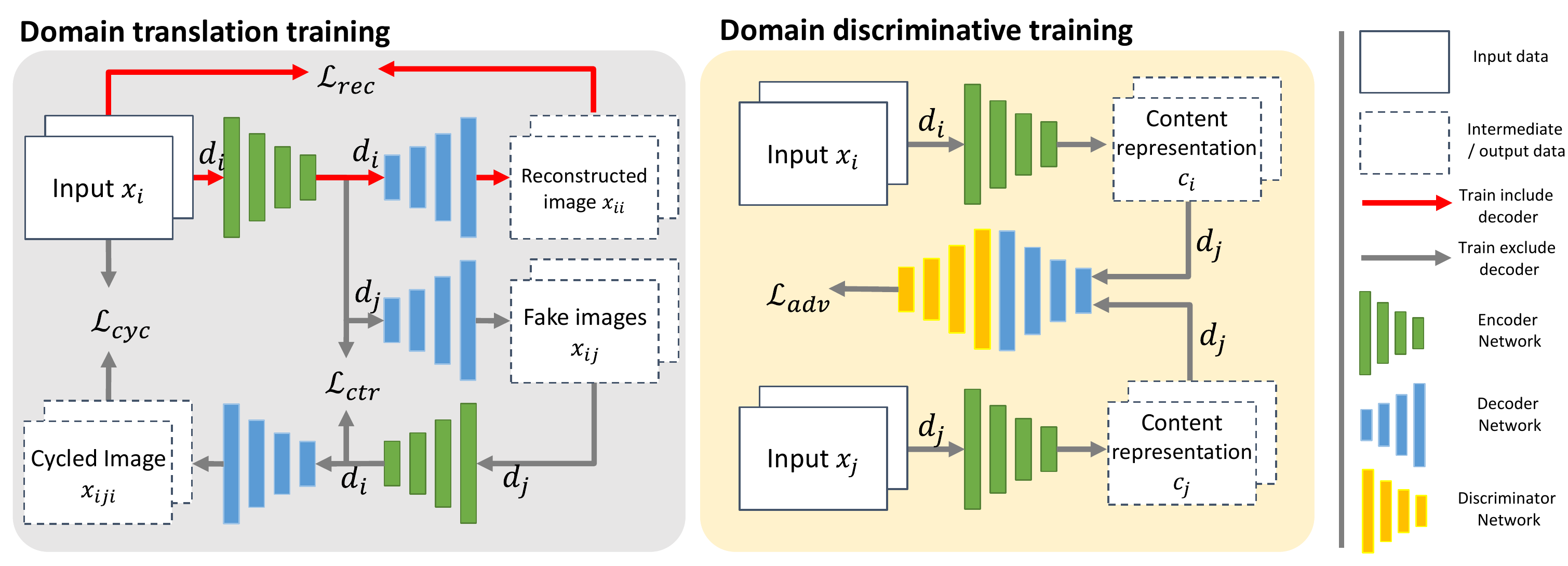}
\caption{Overview of the training scheme in AD-GAN. The grey block shows the within- and cross-domain translation component for Image ${x}_{i}$ in domain $\mathcal{X}_{i}$, while the orange block indicates our domain discriminative training component. ${d}_{i}$ ($i\in \{1,2\}$) denotes the domain label controlling the translation direction.}
\label{fig:scheme}
\end{figure*}

We detail the overall training scheme now. \autoref{fig:scheme} shows an overview of our model training scheme which consists of domain translation training and domain discriminative training. Same- and cross-domain translations are operated in domain translation training to separate off responsibilities of encoder and decoder and establish the relationship between the domain labels and the transferring directions. In domain discriminative training, encoder and decoder are trained to fool a discriminator which in turn tries to differentiate between generated samples and reconstructed samples.

Since content representation $c_1$ and $c_2$ may not be aligned in the shared content space $\mathcal{C}$, which may deteriorate the lossy transformation problem, we further propose a training algorithm to align disentangling representation in the shared space. This algorithm contains two portions: 1) the decoder is only trained during the auto-encoder training; and 2) the discriminator is trained to distinguish between the auto-encoding reconstructed images and generated images. Since $\gend$ is just trained during the auto-encoder training, it can only decode the ${c}_{i} \in \mathcal{C}$ conditioned on the corresponding domain label ${d}_{i}$, which can be treated as a static content reconstruction function during cross-domain translation. As a consequence, the encoder is optimized to fool the discriminator by aligning $c_{i}$ and $c_{j}$ in the latent space, so that ${c}_{j}$ ($j\neq i$) can be decoded conditioned on ${d}_{i}$ for the cross-domain translation.

\begin{figure}[t]
\centering
\includegraphics[width=0.47\columnwidth]{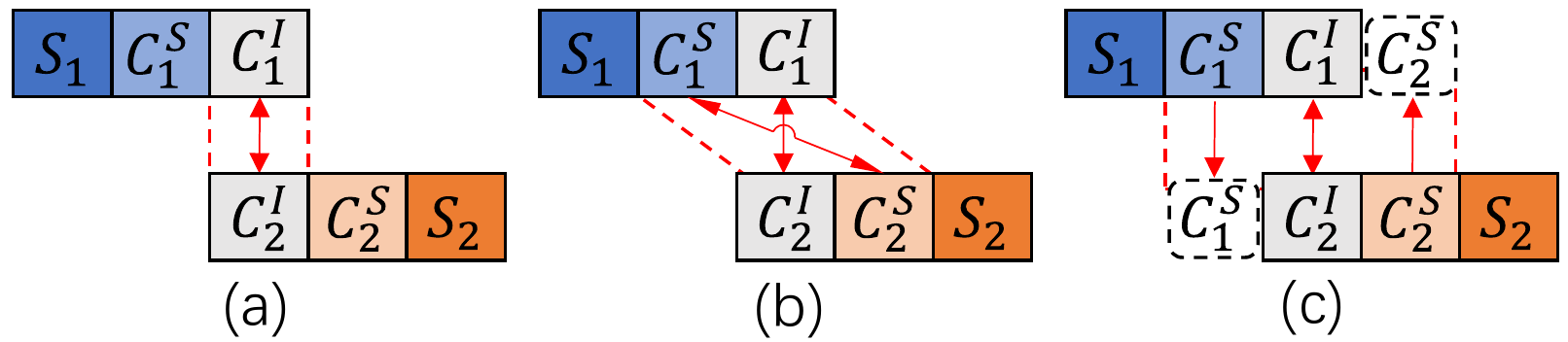}
\caption{Illustration of content and style translation in (a) Disentangled methods, (b) AD-GAN w/o ADT and (c) AD-GAN.}
\label{fig:com}
\end{figure}

\subsection{Further Interpretation}
We present more insight about the proposed AD-GAN. In unpaired image-to-image translation, the exact global properties in the image domain $\mathcal{X}_{1}$ (\eg, nuclei size and amount) are not accessible without ground truth, while the global properties are required to synthesize mask domain $\mathcal{X}_{2}$; this could result in the content inconsistency between image domain $\mathcal{X}_{1}$ and mask domain $\mathcal{X}_{2}$. On the other hand, generating the mask domain with ellipse/ellipsoid simulations could be too simple to reflect the real nuclei shapes, which also leads to content inconsistency when learning cross-domain translations. This is the main challenge in the current unpaired image-to-image translation.  To better explain,  we show an additional illustration in \autoref{fig:com} where each domain $\mathcal{X}_{1}$ or $\mathcal{X}_{2}$ contains three parts: 1) domain-invariant content features $C^I$, \ie the content consistency, 2) domain-specific content features $C^S$, \ie the content inconsistency caused by different distribution, size, and/or number of nuclei of two unpaired domains 3) styles ($S$). \textit{Ideally, we should map $C^I_1$ with $C^I_2$, retain $C_S$ without translating $C^S_1$ and $C^S_2$, and disentangle $S$ from $C$}.

Recent disentangled representation methods \eg MUNIT and DRIT could well learn domain-invariant content features, \ie $C^I_1$ and $C^I_2$. Unfortunately, it does not constrain the styles and assumes  style features sit in a space (see Fig.2b), which cannot reflect the inherit nature of nuclei cell images. A too flexible style space would lead that domain-specific content features $C^S$ will be considered as styles. Moreover, disentangled methods destyle the input images firstly, and then sample a new style to generate cross-domain images. Consequently, $C^S$ would be lost during  cross-domain image translation, resulting in the well-known lossy transformation problem. This can be seen in \autoref{fig:com} (a) where MUNIT aligns $C^I_1$ and $C^I_2$ only while considering both $S_1,S_2$ and $C_2^S, C_1^S$ as styles.

In comparison, AD-GAN makes two major contributions: 1) \textit{a within-domain single style is  assumed} that typically holds in nuclei cell  images. Thus, only $S_1$ and $S_2$ are considered as styles and disentangled. This alleviates the macro-level lossy transformation problem by preserving content explicitly, but they may translate $\{ C_1^S, C_1^I\}$ with $\{ C_2^S, C_2^I\}$ (see \autoref{fig:com} (b)). This still leads to problems, since $C_1^S$ and $C_2^S$ are domain-specific content and should not be translated.  2) \textit{We further propose the novel ADT algorithm} that promotes to align $C_1^I$ with $C_2^I$, but retain $C_1^S$ (or $C_2^S$)   in the other domain (see \autoref{fig:com} (c)). As such, the micro-level lossy transformation problem can be well solved. These contributions lead to substantial performance improvement over the present methods such as MUNIT.

\subsection{Overall Loss}
A one-to-one mapping between the images ${x}_{i}$ and the corresponding content representations ${c}_{i}$ can be built through training an auto-encoder. This can be achieved by the same-domain translation with the image reconstruction loss, which ensures that the generator can reconstruct the original image within a domain. The image reconstruction loss is shown as follows:
\begin{equation}
\small
\mathcal{L}_{rec}(\gene,\gend) = \mathbb{E}_{x_{i}\sim p(x_{i})}[{\left\|\gend(\gene(x_{i},d_{i}),d_{i})-x_{i}  \right \|}_{1}].
\label{eqn:1}
\end{equation}

GAN is typically used in order to build a one-to-one mapping between $c \in \mathcal{C}$ and two image domains $\mathcal{X}_{1} $ and $\mathcal{X}_{2}$ in unsupervised content and style disentanglement~\cite{MUNIT,DRIT,DRIT_plus}. In our adversarial training, the discriminator is trained to distinguish the reconstructed and fake images, which generated by the frozen decoder $\gend^*$. The adversarial loss is shown as follows:
\begin{equation}
\small
\begin{split}
&\mathcal{L}_{adv}(\gene, \dis) =\\
& \mathbb{E}_{x_{i} \in p(x_{i})}[\log \dis(\gend^*(\gene(x_{i},d_{i}),d_{i})|d_{i})] \\
&+ \mathbb{E}_{x_{j} \in p(x_{j})}[1-\log \dis(\gend^*(\gene(x_{j},d_{j}),d_{i})|d_{i})],
\end{split}
\label{eqn:2}
\end{equation}
Here under the aligned disentangling training, since $\gend^*$ is a conditional content reconstruction function, the discriminator $\dis$ is actually trained to distinguish from $c_i$ and $c_j$, where $c \in \mathcal{C}$ is the encoded representation by $\gene$, $i \neq j$. As a consequence, $c_i$ and $c_j$ shall be aligned by the generator on semantic object level to ensure $c_j$ can be decoded conditioned on $d_i$ to fool the discriminator.

We also exploit the content reconstruction loss that proves still useful even after we apply the aligned disentangling training. This loss prevent the content representation changed in across domain, which is expressed as follows:

\begin{equation}
\small
\begin{split}
&\mathcal{L}_{ctr}(\gene)=\\
&\mathbb{E}_{x_{i} \in p(x_{i})}[{\left \|\gene(\gend^*(\gene(x_{i},d_{i}),d_{j}),d_{j})-\gene(x_{i},d_{i})  \right \|}_{1}].
\end{split}
\label{eqn:3}
\end{equation}

\begin{table*}[htb]
\centering
\caption{Quantitative comparison on 2D data. Here ${}^{\dagger}$ indicates a supervised segmentation model and ${}^{*}$ indicates a two-stage model trained with paired synthetic data.}
\begin{tabular}{|c|c|c|c|c|c|c|c|}
\hline
\multirow{2}{*}{}             & \multirow{2}{*}{Methods} & \multicolumn{3}{c|}{Fluo-N2DL-HeLa}                                   & \multicolumn{3}{c|}{HaCaT}                                            \\ \cline{3-8}
                              &                          & Precision             & Recall                & DICE                  & Precision             & Recall                & DICE                  \\ \hline\hline
\multirow{6}{*}{Unsupervised} & CycleGAN                 &74.3$\pm$10.2          & 70.6$\pm$11.8         & 72.2$\pm$10.3          & 75.2$\pm$13.7         & 58.3$\pm$16.0         & 63.6$\pm$13.7         \\ \cline{2-8}
                              & MUNIT &81.8$\pm$12.7   &  63.5$\pm$9.9  &71.2$\pm$11.0 &  78.4$\pm$8.1   & 54.6$\pm$16.3     &  61.6$\pm$12.7    \\ \cline{2-8}
                              & UDCT                     & 85.2$\pm$3.5          & 82.4$\pm$3.8          & 83.7$\pm$3.3          & 77.1$\pm$9.6          & 66.1$\pm$10.5         & 69.5$\pm$8.81         \\ \cline{2-8}
                              & CGU-net${}^{*}$  &72.3$\pm$2.1&79.6$\pm$2.6 &75.7$\pm$0.9  & 83.6$\pm$0.4 & 66.8$\pm$0.7 & 73.5$\pm$0.4 \\\cline{2-8}
                              & UFDN  &    80.9$\pm$0.5       &      73.9$\pm$0.8     &    77.0$\pm$0.2   &      88.8$\pm$0.7      &    71.0$\pm$1.3     &  77.8$\pm$0.6   \\ \cline{2-8}
                              & AD-GAN                    & \textbf{92.8$\pm$1.0} & \textbf{89.2$\pm$1.4} & \textbf{90.9$\pm$0.6} & \textbf{85.4$\pm$0.5} & \textbf{95.2$\pm$0.1} & \textbf{89.3$\pm$0.2} \\ \hline
\multirow{2}{*}{Supervised}   & U-Net${}^{\dagger}$                    & 88.4$\pm$2.1          & 93.9$\pm$0.8          & 91.0$\pm$1.4          & 89.1$\pm$3.3          & 93.2$\pm$3.2          & 90.5$\pm$0.8          \\ \cline{2-8}
                              & nnU-Net${}^{\dagger}$                  & 97.1$\pm$0.3          & 83.1$\pm$0.2          & 89.1$\pm$0.2          & 87.8$\pm$0.4          & 98.2$\pm$0.2          & 92.7$\pm$0.3   \\ \hline
\end{tabular}
\label{tab:2ddata}
\end{table*}
\begin{figure*}[htb]
\centering
\includegraphics[width=1\columnwidth]{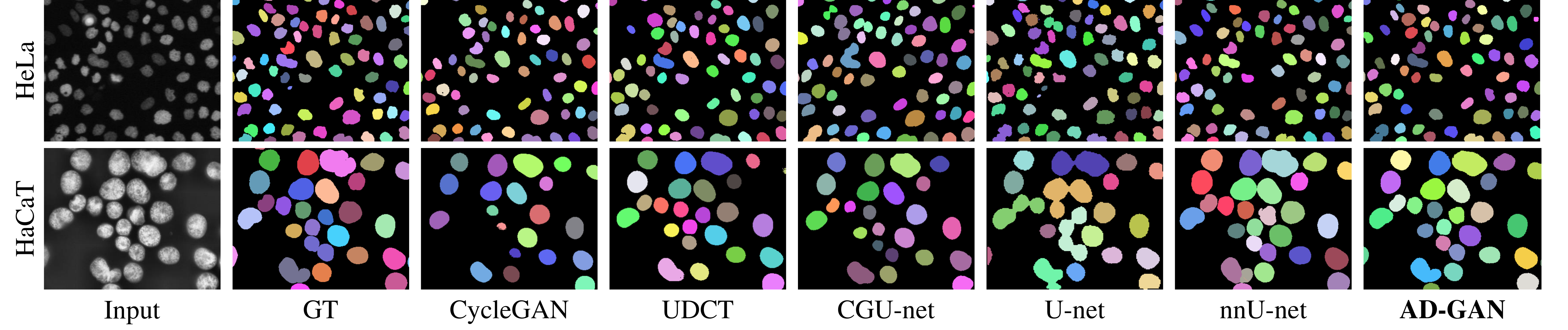}
\caption{Visualization of semantic segmentation results on 2D data Fluo-N2DL-HeLa and HaCaT.}
\label{fig:hacat}
\end{figure*}

Moreover, we utilize the cycle consistency loss proposed in CycleGAN~\cite{CycleGAN}. When a given input ${x}_{1}$ passes through the cross-domain translation pipeline $\mathcal{X}_{1} \rightarrow \mathcal{X}_{2} \rightarrow \mathcal{X}_{1}$, it should be able to be reconstructed back to ${x}_{1}$ itself. The cycle consistency loss is shown as follows:

\begin{equation}
\small
\begin{split}
&\mathcal{L}_{cyc}(\gene)= \\
&\mathbb{E}_{x_{i}\in p(x_{i})}[{\left \|\gend^*(\gene(\gend^*(\gene(x_{i},d_{i}),d_{j}),d_{j}),d_{i})-x_{i}  \right \|}_{1}].
\end{split}
\label{eqn:4}
\end{equation}

Finally, we jointly  train the  encoder, decoder, and domain discriminator to optimize the full objective:
\begin{equation}
\small
\nonumber
\arg\mathop{\min}_{\gene,\gend}\arg\mathop{\max}_{\dis} \mathcal{L}_{adv}+\lambda_{cyc}\mathcal{L}_{cyc}
+\lambda_{rec}\mathcal{L}_{rec}+\lambda_{ctr}\mathcal{L}_{ctr},
\end{equation}
where each term is respectively described in \autoref{eqn:1}-\ref{eqn:4}, and $\lambda_{cyc}$, $\lambda_{rec}$, $\lambda_{ctr}$ are the trade-off parameters to adjust the importance of each term.
%where each term is respectively described in \eqref{eqn:1} and $\lambda_{cyc}$, $\lambda_{rec}$, $\lambda_{ctr}$ are the trade-off parameters to adjust the importance of each term. %are the trade-off parameters to
\section{Experiments}
\label{sec:experimental-results}

\subsection{Datasets}
\noindent\textbf{Fluo-N2DL-HeLa}  is a 2D  benchmark set from the cell tracking challenge~\cite{Neumann2010}, containing 2 sequences of labelled and 2 sequences of unlabeled images.  For unsupervised models, we use all images without any labels for training. For supervised models, we use sequence 01 to train the model. All the methods are evaluated on sequence 02. \\
\textbf{HaCaT} is 2D dataset collected from a recently published dataset S-BSST265~\cite{CaHaT}. It contains 26 training images and 15 testing images with different magnification, which are scaled to the same magnification (20$\times$).\\
\textbf{BBBC024}  is a 3D dataset~\cite{bbbc024} containing 80 simulated HL60 cell nuclei images with different degree of clustering collected from Board Bioimage Benchmark Collection.\\
\textbf{Scaffold-A549} is a 3D fluorescence microscopy dataset for A549 cell culture on bio-scaffold, which contains 20 unlabelled images for training and one image was fully annotated for quantitative measurement.

\subsection{Implementation And Training Details}
We implemented our framework with the open source software library PyTorch 1.8.0 on a workstation equipped with one NVIDIA GeForce RTX 2080 Ti GPU. We follow~\cite{MUNIT} and adopt the its architecture for our generative networks, see the appendix for details. For 2D datasets,
cropped patches with size of $256\times256$ are ustilized to train the model under the mini-batch of 16. For 3D datasets, we replace the 2D convolutional layers with 3D version, and halve the channel number for memory saving. Then the cropped sub-volumes with size of $64\times128\times128$ are used to train the model with batch size of 4. Adam~\cite{adam} has been used as an optimizer to minimize objective function an initial learning rate of 0.0001 and weight decay of 0.0001. The learning rate remains unchanged for the first 5000 iterations and is linearly decayed  to zero over the next 5000 iterations. Basic data augmentation are engaged to avoid overfitting, including flipping, rotation and random crop. In all the experiments, we tune the weight $\lambda_{ctr}=1$, $\lambda_{cyc}=20$ and $\lambda_{rec}=20$ empirically.
% with basic data
% the model is trained for 10000 iterations, and the

% we tune the weight $\lambda_{ctr}$, $\lambda_{cyc}$ and $\lambda_{rec}$ empirically.
% All the models are trained for 10000 iterations

% all the model are train the model for 10000 iterations
% % \noindent\textbf{Network Architecture:}
% % For the generator used in AD-GAN, the encoder consists of 2 down-sampling blocks and 4 residual blocks, while the decoder exploits 4 residual blocks and 2 up-sampling blocks, which all equipped with AdaIN~\cite{adain}. The discriminator was designed based on PatchGAN~\cite{PatchGAN}, which contains LeakyReLU for nonlinearity.\\

% \textbf{Training Strategy:} We follow the setting in CycleGAN~\cite{CycleGAN} and use LSGANs~\cite{lsgan} to stabilize the training. In all the experiments, we tune the weight $\lambda_{ctr}$, $\lambda_{cyc}$ and $\lambda_{rec}$ empirically. The ADAM solver~\cite{adam} is used with a batch size of 16. All the networks are trained from scratch with a learning rate of $0.0001$ and weight decay $0.0001$. The learning rate remains unchanged for the first 100 epochs and is linearly decayed  to zero over the next 100 epochs.

\subsection{Results on 2D Datasets}

We first compare our method with competitive models in unsupervised nuclei segmentation on two 2D benchmark datasets. In particular, CGU-net~\cite{Bohland2019Influ} and UDCT~\cite{Ihle2019UDCT} are two latest methods which probably obtain the best performance so far in 2D unsupervised nuclei segmentation task. We select CycleGAN~\cite{CycleGAN} and MUNIT~\cite{MUNIT}, two most representative general-purpose unpaired image-to-image translation under different assumptions (see \autoref{fig:comparison}). We also
compare the model of UFDN~\cite{UFDN} which has the similar assumption with ours but utilizes a VAE-based disentanglement.

Following the previous research, we take the pixel-based metric to evaluate the performance. Specifically, we report  precision, recall, and DICE coefficient~\cite{vnetdice} to evaluate different approaches. Here DICE = $\frac{2 \times n_{TP}}{n_{TP} + n_{FP}+n_{TP} + n_{FN}}$, where $n_{TP}$, $n_{FP}$, $n_{FN}$ are defined to be the number of true-positives, false-positives and false-negative segmentation result pixels in an image, respectively. A higher DICE coefficient indicates a better intersection between the ground truth and the predicted segmentation masks.

\begin{figure*}[htb]
\centering
\includegraphics[width=0.9\columnwidth]{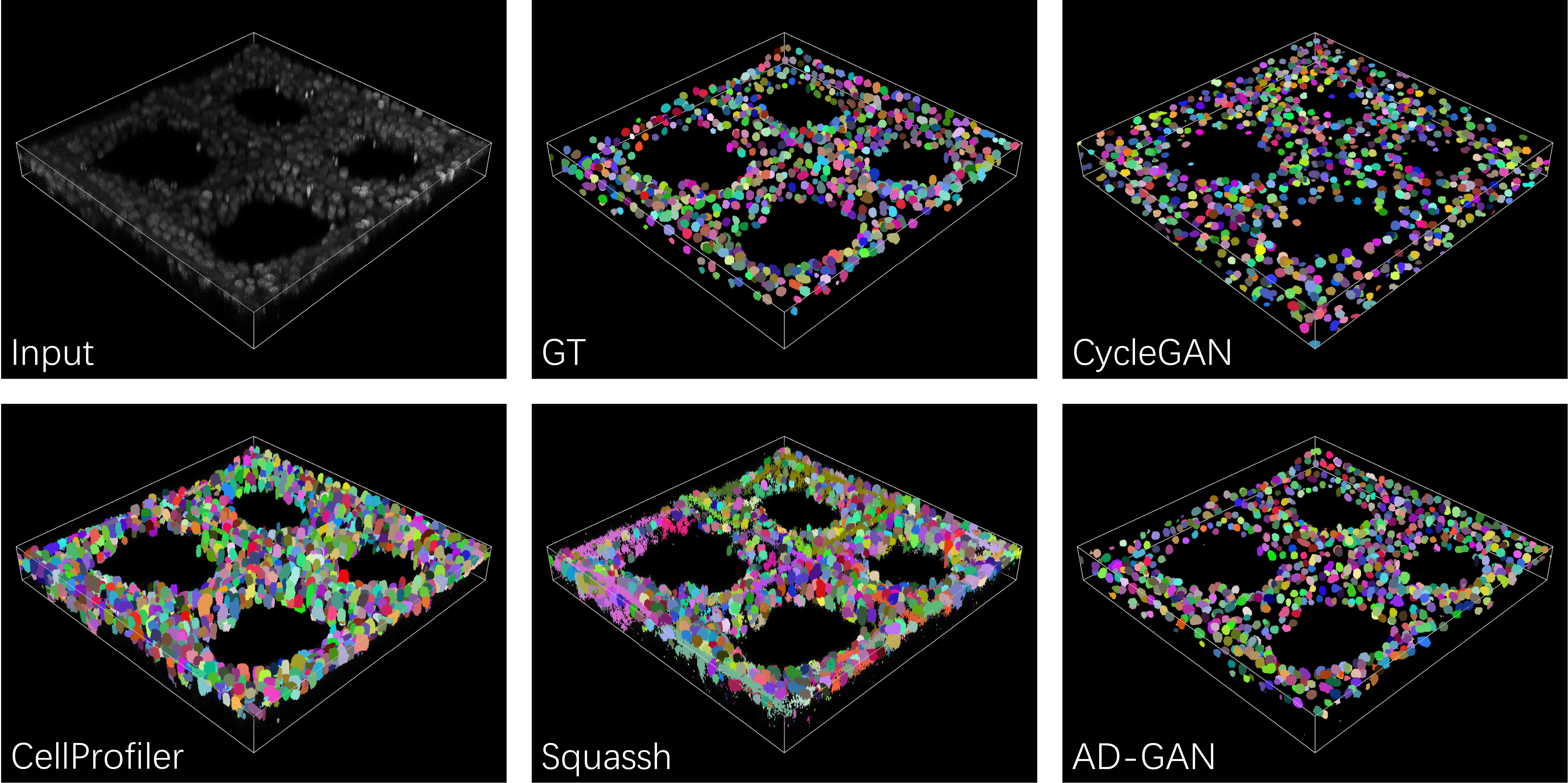}
\caption{Visualization of semantic segmentation results on 3D data scaffold-A549. }
\label{fig:3ddata}
\end{figure*}

We train each method for five times and report the means and standard deviation for quantitative evaluations. The comparison results are shown in \autoref{tab:2ddata}. As observed, our proposed AD-GAN achieves significantly higher performance in DICE than all the unsupervised models on both the 2D datasets. More specifically, AD-GAN attains $90.9\%$ and $89.3\%$  on Fluo-N2DL-Hela and HacaT respectively, which improve relatively the best of the other models by $8.3\%$ and $14.8\%$. In addition, it is noted that GAN-based methods \eg CycleGAN, MUNIT, and UDCT, generate a much larger standard deviation compared with that of AD-GAN; meanwhile, the two-stage method CGU-net enjoys a low standard deviation, while its performance highly depends on the quality of synthetic data. On the other hand, although the VAE-based method UFDN achieves good results thanks to its assumption similar  to us, UFDN struggles to predict  with the high recall due to its partial disentanglement, thereby resulting in a much lower dice than our proposed AD-GAN.

% In addition, it is noted that GAN-based methods \eg CycleGAN, MUNIT and UDCT, generate a much larger standard deviation compared with that of AD-GAN. \textcolor{orange}{Despite high precision, VAE-based method UFDN struggles to predict results with high recall due to the partial disentanglement.} On the other hand, although two-stage method CGU-net enjoys a low standard deviation as well, its performance highly depends on the quality of synthetic data. Stable performance presents one appealing feature of our AD-GAN model.

To further examine the AD-GAN's performance, we even compare it with two famous supervised models, \ie U-Net~\cite{unet} and nnU-Net~\cite{nnunet}.  nnU-Net is one of the state-of-the-art supervised semantic segmentation method on many biomedical data, which is a standardized baseline with no need for the manual intervention. As seen in the bottom part of \autoref{tab:2ddata}, despite its unsupervised nature, AD-GAN can surprisingly lead to very similar performance to the supervised models. Without  expensive manual annotations, the unsupervised AD-GAN model has a great potential to be applied in practice.

Additionally, we engage a simple post-processing containing morphological erosion and 2D watershed to visualize the segmentation results. We illustrate in \autoref{fig:hacat} one example for each of the two 2D dataset respectively. As clearly observed, those CycleGAN based methods, \ie the standard CycleGAN, UDCT, CGU-net,  typically lead to the the lossy transformation problem, \eg nuclei offset, shape inconsistency, nuclei deletion or addition. In comparison, our proposed AD-GAN can reduce such negative effects significantly and shows much better performance among all the unsupervised models.

Finally, we also report more visualization results in appendix, which further shows the advantages of our proposed method.

\begin{table}[t]
\centering
\caption{Comparison on 3D data BBBC024 and Scaffold-A549. Here ${}^{\dagger}$ indicates a supervised segmentation model and ${}^{*}$ indicates a two-stage model trained with paired synthetic data.}
\begin{tabular}{|c|c|ccc|}
\hline
                              & Methods           & Precision     & Recall        & DICE          \\ \hline\hline
\multirow{7}{*}{Unsupervised} &                   & \multicolumn{3}{c|}{BBBC024}                  \\
                              & CellProfiler      & 81.3          & 91.0          & 85.9          \\
                              & ImageJ Squassh    & 76.1          & \textbf{99.5} & 86.2          \\
                              & CycleGAN          & \textbf{94.0} & 63.4          & 75.8          \\
                              & SpCycleGAN        & 76.4          & 71.1          & 73.7          \\
                              & DeepSynth${}^{*}$ & 63.4          & 66.6          & 64.9          \\
                              & AD-GAN            & 93.8          & 91.5          & \textbf{92.6} \\ \hline
\multirow{2}{*}{Supervised}   & U-Net${}^{\dagger}$              & 94.2             & 94.8             & 94.5             \\
                              & nnU-Net${}^{\dagger}$            & 95.1             &95.3             & 95.2             \\ \hline\hline
\multirow{7}{*}{Unsupervised} &                   & \multicolumn{3}{c|}{Scaffold-A549}            \\
                              & CellProfiler      & 37.5          & \textbf{91.3} & 53.1          \\
                              & ImageJ Squassh    & 51.7          & 78.6          & 62.3          \\
                              & CycleGAN          & 53.9          & 41.7          & 47.0          \\
                              & SpCycleGAN        & 52.3          & 47.7          & 49.7          \\
                              & DeepSynth${}^{*}$ & 49.9          & 42.3          & 45.8          \\
                              & AD-GAN            & \textbf{89.0} & 78.2          & \textbf{83.3} \\ \hline
\end{tabular}
\label{tab:3ddata}
\end{table}

\subsection{Results on 3D Datasets}

\begin{table*}[htp]
\centering
\caption{Ablation study of AD-GAN. Note that on Fluo-N2DL-HeLa, the performance is measured on both the sequences.}
\begin{tabular}{|c|cccccc|}
\hline
Same-domain image reconstruction $\mathcal{L}_{rec}$ &- & \checkmark& \checkmark& \checkmark & \checkmark  & \checkmark   \\
Cross-domain content reconstruction $\mathcal{L}_{ctr}$&\checkmark&- & \checkmark&  \checkmark& \checkmark   & \checkmark  \\
Cycle Consistency $\mathcal{L}_{cyc}$&\checkmark&\checkmark&-&  \checkmark & \checkmark  & \checkmark\\
AdaIN in Encoder&\checkmark&\checkmark&\checkmark& -& \checkmark  & \checkmark \\
Aligned disentangling training   & -   &      -     &    -       &  -&-& \checkmark\\ \hline\hline
Fluo-N2DL-HeLa          &87.9  & 85.1&64.4&88.0 &89.1 &\textbf{91.1}\\
HaCaT                   &78.9  &78.7& 49.2&79.6 & 80.2&\textbf{89.3}\\
BBBC024                 &76.5  &75.8 & 58.0& 79.2& 80.6&\textbf{92.6}\\
Scaffold-A549           & 75.6 &73.2&52.4& 78.3&80.2 &\textbf{83.3}
  \\ \hline
\end{tabular}
\label{tab:ablation2}
\end{table*}

\begin{table}[t]
\centering
\caption{Comparison of AD-GAN-INS and the state-of-the-art top three methods on Cell Tracking Challenge benchmark. 'First', 'Second' and 'Third' represent the top-three methods on the leader board.}
\begin{tabular}{|c|ccc|}
\hline
Methods            & DET        & SEG       & $\rm{OP}_{\rm{csb}}  $                          \\ \hline\hline
                  & \multicolumn{3}{c|}{Fluo-N2DL-HeLa}                                                                      \\
First                & 0.994      & 0.923     & 0.957     \\
Second               & 0.992      & 0.923     & 0.954     \\
Third                & 0.992      & 0.919     & 0.953     \\
CycleGAN             & 0.823      & 0.486     & 0.654     \\
AD-GAN-INS           & 0.938      & 0.850     & 0.894     \\ \hline\hline
                  & \multicolumn{3}{c|}{Fluo-N3DH-CHO}                                                                \\
First                & 0.954      & 0.917     & 0.926     \\
Second               & 0.945      & 0.914     & 0.913     \\
Third                & 0.934      & 0.903     & 0.913     \\
CycleGAN             & 0.782      & 0.506     & 0.644     \\
AD-GAN-INS           & 0.881      & 0.823     & 0.852     \\ \hline
\end{tabular}
\label{tab:ctc}
\end{table}

We now evaluate the various models on the more challenging 3D data for unsupervised nuclei segmentation. For fair comparison, we follow SpCycleGAN~\cite{Fu2018three} and use the original CycleGAN  as the baseline to synthesize the image masks. To adapt to 3D image-to-image translation task, we replace the 2D convolution layer with the 3D convolution layer and decrease the dimension of each layer to half for memory saving. We use the typical voxel-based metric to compare our method AD-GAN with the existing competitive models in unsupervised 3D nuclei segmentation including the original CycleGAN~\cite{CycleGAN}, SpCycleGAN~\cite{Fu2018three}, and  the DeepSynth~\cite{DeepSynth}. Particularly, DeepSynth is widely recognized as the state-of-the-art deep model for unsupervised 3D nuclei segmentation. We evaluate our model against the  famous biomedical image processing tools CellProfiler 3.0~\cite{cellprofiler} and Squassh~\cite{squassh}. We do not compare UFDN, UDCT, CGU-net*, and MUNIT in the 3D scenario since 1) they are not originally designed for 3D segmentation, and 2) there are no source codes available  for these models in the 3D case. The baseline supervised methods are only applied on the BBBC024 dataset since no annotation is available in Scaffold-A549 dataset.

The voxel-based segmentation results are again evaluated quantitatively based on precision, recall and DICE. These results are reported in \autoref{tab:3ddata}. As observed on both the 3D datasets, biomedical image processing tools tend to recognize the background as foreground; thus they could over-segment the images,  resulting in a high recall but low precision.
%As observed on both 3D datasets, biomedical image processing tools tend to recognize the background as foreground so that over-segment the images, which results in a high recall but low precision.
On BBBC024 dataset which contains relatively simple data, CycleGAN based methods can obtain encouraging results. However, a position offset and shape difference problem  can still be observed in the experiments. In comparison, our proposed AD-GAN shows the very promising result of $92.6\%$, which is much higher than all the other unsupervised approaches. Meanwhile, Scaffold-A549 is a more challenging dataset where cells tend to grow on the scaffold, making the object distribution non-uniform and highly complicated.  Uneven distribution of nuclei in 3D images makes the lossy transformation problem even worse. Therefore, a large content inconsistency would exist between the real image data and synthetic mask data. All the other comparative models perform poorly on this dataset. In contrast, our proposed AD-GAN fully utilizes the advantages of the end-to-end training and obtains $83.3\%$ on Scaffold-A549 w.r.t. DICE, which is substantially higher than the best of the comparison models, achieved by Squassh. Finally, we also visualize the segmentation results of the various approaches in \autoref{fig:3ddata}. It can be clearly observed that our proposed AD-GAN model leads to much better segmentation performance.
More visualization results can also be seen in appendix, which again shows the advantages of our proposed method.

\begin{figure}
\centering
\includegraphics[width=0.47\columnwidth]{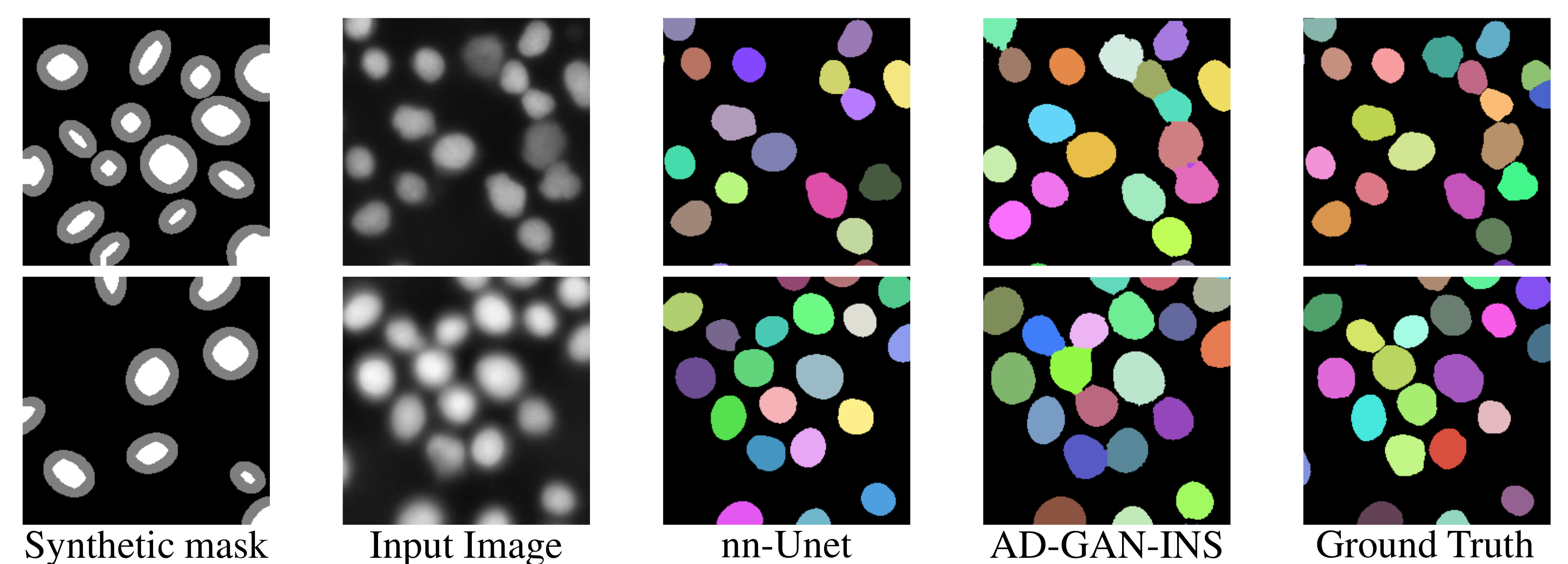}
\caption{Visualization of instance segmentation on HaCaT}
\label{fig:ins}
\end{figure}

\subsection{Instance Segmentation}
By better coping with the lossy transformation problem, our method can be readily extended to instance segmentation. Without modifying other training settings, we exploit the gray color to represent the edge of each object and use them to generate the mask domain. Once such a model is trained (called as AD-GAN-INS), the instance segmentation can be obtained with a threshold-based image ternarization and marker-based watershed algorithm, as shown in the fourth column of \autoref{fig:ins}. Compared with the AD-GAN semantic segmentation outputs, clustered nuclei can be separated efficiently. Particularly, on HaCaT we got the object-based F1-Score (0.5 IoU threshold) of 95.2\%, which is competitive when compared with 97.3\% obtained by the supervised method nnUnet\footnote{The instance segmentation results were obtained by three classes semantic segmentation (foreground, boundary, and background) and watershed algorithm post-processing.}.

To further evaluate our method with the state-of-the-art, we conduct instance segmentation on Fluo-N2DL-HeLa and Fluo-N3DH-CHO within Cell Tracking Challenge\footnote{Cell Tracking Challenge page: \href{http://celltrackingchallenge.net}{http://celltrackingchallenge.net}}.  We follow the official metrics to report the performance by SEG, DET, and $\rm{OP}_{\rm{csb}}$, where $\rm{OP}_{\rm{csb}}= 0.5\times(\rm{SEG}+\rm{DET})$. As shown in \autoref{tab:ctc}, we improve the unsupervised baseline CycleGAN by a large margin, especially on segmentation sub-task. Meanwhile, the overall performance of our AD-GAN-INS is closed with the state-of-the-art without supervision or hyperparameter tuning. Further works allow us to further minimize the gap between supervised and unsupervised methods. See the appendix for more details.

% the lossy transformation problem significantly degrade the performance of segmentation for CycleGAN, results in a 0.486 and 0.506 scores on the two datasets, respectively. Meanwhile, the nuclei addition/deletion also

%\footnote{\href{http://celltrackingchallenge.net/files/leaderboards/CSB/2021-10-22.png}{Cell segmentation benchmarck on 2021-10-22.}}

\subsection{Image Synthesis}

\begin{figure}
\centering
\includegraphics[width=0.47\columnwidth]{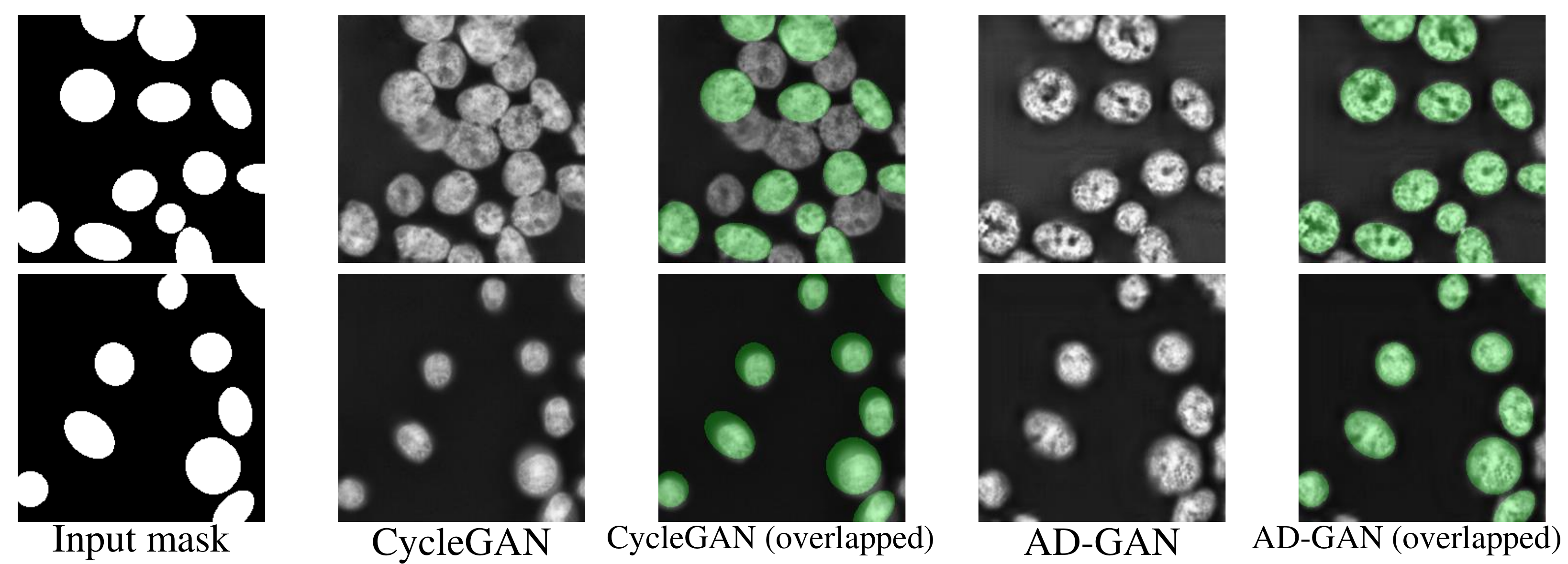}
\caption{Visualization of image synthesis results on HaCaT.}
\label{fig:synthesis}
\end{figure}

Although our method can output instance segmentation end-to-end in an unsupervised manner, it is required in more complex tasks, \eg cell tracking, to generate or synthesize training data. In such case, the performance of generating cell images from synthetic masks is also essential. We demonstrate that AD-GAN could lead to much better performance than the standard CycleGAN when generating cell images from the mask domain in \autoref{fig:synthesis}. To better visualize the difference (or highlight the lossy transformation problem in CycleGAN), we also show the input masks and generated images together in one single image (see the 3rd and 5th column of \autoref{fig:synthesis}). As observed, without any shape difference and position offset, our method can generate more realistic images. Detailed experiments about two-stage pipeline with AD-GAN can be found in the appendix.

% can encourage the model to preserve the content by

% preserve the content during cross-domain translation and learn well-disentangled representations.

% With the component of the same-domain image translation, we observe that the content position offset problem can be significantly alleviated.
% With the content reconstruction component, it is observed that the content deletion/addition problem can be much alleviated, which also improves the DICE score.
% However, the DICE score significantly drops without the cycle consistency component which is an explicit image-level constraint term in image-to-image translation.

% The ablation study on the three components suggests that $\mathcal{L}_{rec}$, $\mathcal{L}_{ctr}$ and $\mathcal{L}_{cyc}$ can encourage the model to preserve the content during cross-domain translation and learn well-disentangled representations.

\subsection{Ablation Study}
\begin{figure}[tp]
\centering
\includegraphics[width=0.47\columnwidth]{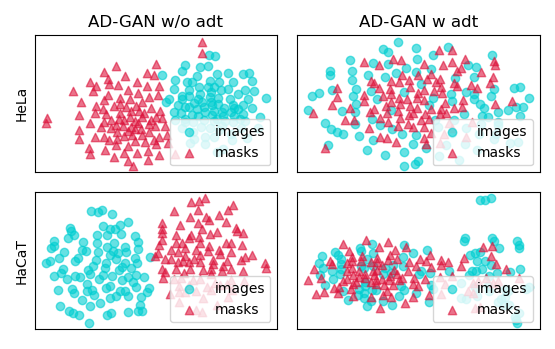}
\caption{Visualization of content representations of two domains using t-SNE~\cite{2008Visualizing}. adt denotes aligned disentangling training.}
\label{fig:ablationtsne}
\end{figure}

To analyze the importance of different components in our model, we conduct an ablation study with five variants of AD-GAN, as shown in \autoref{tab:ablation2}.
The ablation study on three components suggests that $\mathcal{L}_{rec}$, $\mathcal{L}_{ctr}$ and $\mathcal{L}_{cyc}$ suggest that the reconstruction on image-level and feature-level are important to establish correspondence between content and style. Once built, it is observed that the macro-level lossy transformation problem have been alleviated by preserving content explicitly during cross-domain translation. We also find that adding AdaIN in the encoder also helps improve the performance, since introducing domain information determines the translation direction. Finally, when we implement the aligned disentangling training, AD-GAN can enforce the disentangled content for each domain  to be well aligned in the latent space as visualized in  \autoref{fig:ablationtsne}. This consequently benefits largely the content preserving and helps reduce the  micro-level lossy transformation problem.

\subsection{Parameter Sensitivity Analysis}
\begin{figure}[tp]
\centering
\includegraphics[width=0.47\columnwidth]{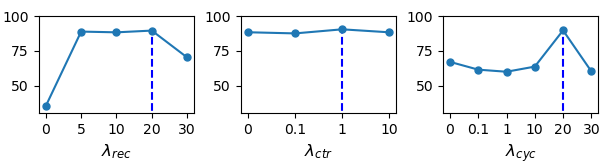}
\caption{Parameter sensitive analysis on Fluo-N2DL-HeLa.}
\label{fig:se}
\end{figure}

We further investigate the parameter sensitivity of the trade-off parameters involved in the loss function, \ie  $\lambda_{ctr}$, $\lambda_{cyc}$, and $\lambda_{rec}$, which are reported sensitivities in \autoref{fig:se}. In most evaluations, we fix $\lambda_{ctr} = 1$. Since gray-scale images are used in our tasks and the weight for image reconstruction need be enlarged to train the decoder during aligned disentangling training, $\lambda_{cyc}$ and $\lambda_{rec}$ are tuned to 20 for all experiments, which is typically 10 in other CycleGAN based methods. It can be seen that $\lambda_{ctr}$ is not sensitive,  $\lambda_{cyc}$ and $\lambda_{rec}$ are suggested to be the same or close.

\subsection{Analysis on the Nuclei Number in Synthetic Masks}
Due to its unsupervised nature, our AD-GAN may ignore nucleus with low brightness and classify them as background in some cases. This can be alleviated by increasing the number of nucleus in synthetic masks during the training phase. As such, AD-GAN tends to identify more objects to be mapped. However, the exact nuclei number is not available for unsupervised nuclei segmentation. In this subsection, we examine if such  nuclei number pre-specified in  synthetic masks would affect the performance of various methods.

\begin{figure}[tp]
\centering
\includegraphics[width=0.47\columnwidth]{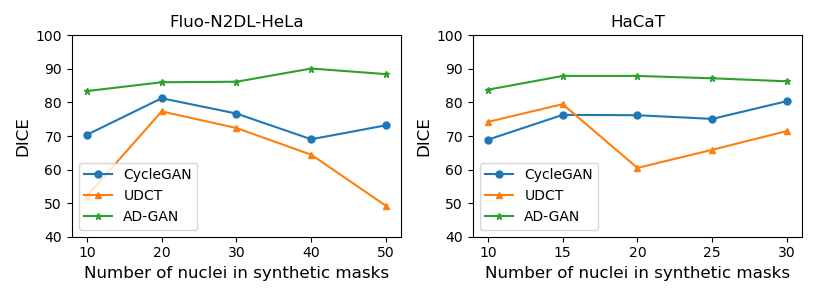}
\caption{Performance comparison vs. number of nuclei cells in synthetic masks.}
\label{fig:nc}
\end{figure}

We conduct some experiments to  evaluate the image-to-mask translation performance with different content property settings for synthesizing masks. Particularly, we train CycleGAN, UDCT, and AD-GAN with the different  nuclei number $n$ when generating the synthetic masks. In another word, the total number of objects in each synthetic mask is controlled in [$n$/2, $n$]. The results in \autoref{fig:nc} show that  CycleGAN and UDCT highly depend on an appropriate selection of nuclei number where the largest gap between best and worst results is even more than 10\%. As a matter of fact, this sensitivity issue of CycleGAN and UDCT has been earlier  discussed in CGU-net~\cite{Bohland2019Influ} and UDCT~\cite{Ihle2019UDCT}. On the other hand, AD-GAN leads to very stable results, which are insensitive to  $n$. This is mainly because our proposed AD-GAN enjoys a loose restriction between the matching of content between the input nuclei images and the synthetic masks. More detailed qualitative comparison on Fluo-N2DL-HeLa can be found in the appendix.

\section{Conclusion and Future Work}
\label{sec:conclusion}
In this paper, we propose the Aligned Disentangling Generative Adversarial Network (AD-GAN) for end-to-end unsupervised nuclei segmentation. AD-GAN takes advantages of disentangled representation under within-domain same-style assumption to reduce macro-level lossy transformation problem.  We further propose a novel training strategy to align the disentangled content representation in the hidden space in order to release micro-level lossy transformation problem. Our proposed method can be readily extended to instance segmentation tasks and image synthesis directly. Compared with existing deep-learning based unsupervised nuclei segmentation methods, AD-GAN demonstrates significantly better performance on both 2D and 3D data. In the future, we will consider to  relax the within-domain same-style assumption and explore the possibility to extend our proposed model on unsupervised image-to-mask translation in  more complicated medical scenarios.

\bibliographystyle{IEEEtran}
\bibliography{refs}

\newpage
\section*{Appendix}
\subsection{Network Architecture}
The architecture details for 2D nu
clei segmentation are reported as follows: $h$ and $w$: height and width of the input images, $n_s$: the dimension of the AdaIN parameters, $N$: the number of output channels, $K$: kernel size, $S$: stride size, $P$: padding size, FC: fully connected layer, IN: instance normalization, AdaIN: adaptive instance normalization, ReLU: rectified linear unit, LReLU: Leaky ReLU with a negative slope of 0.2.

\begin{table*}[ht]
\centering
\caption{Architecture of the unified content encoder $\gene$. The style representation learned from MLP for each domain is injected by adaptive instance normalization.}
\begin{tabular}{ccc}
Part                           & Input $\rightarrow$ Output Shape  & Layer Information                  \\ \hline\hline
\multirow{3}{*}{Down-sampling} & (h,w,1)+(64+64) $\rightarrow$ (h,w,64)                                                    & CONV-($N$64, $K$7x7, $S$1, $P$3), AdaIN, ReLU            \\
                               & (h,w,64)+(128+128) $\rightarrow$ ($\frac{h}{2}$,$\frac{w}{2}$,128)                          & CONV-($N$128, $K$3x3, $S$2, $P$1), AdaIN, ReLU           \\
                               & ($\frac{h}{2}$,$\frac{w}{2}$,128)+(256+256) $\rightarrow$ ($\frac{h}{4}$,$\frac{w}{4}$,256) & CONV-($N$256, $K$3x3, $S$2, $P$1), AdaIN, ReLU           \\ \hline
\multirow{4}{*}{Bottleneck}    & ($\frac{h}{4}$,$\frac{w}{4}$,256)+(256+256) $\rightarrow$ ($\frac{h}{4}$,$\frac{w}{4}$,256) & ResBlock: CONV-($N$256, $K$3x3, $S$1, $P$1), AdaIN, ReLU \\
                               & ($\frac{h}{4}$,$\frac{w}{4}$,256)+(256+256) $\rightarrow$ ($\frac{h}{4}$,$\frac{w}{4}$,256) & ResBlock: CONV-($N$256, $K$3x3, $S$1, $P$1), AdaIN, ReLU \\
                               & ($\frac{h}{4}$,$\frac{w}{4}$,256)+(256+256) $\rightarrow$ ($\frac{h}{4}$,$\frac{w}{4}$,256) & ResBlock: CONV-($N$256, $K$3x3, $S$1, $P$1), AdaIN, ReLU \\
                               & ($\frac{h}{4}$,$\frac{w}{4}$,256)+(256+256) $\rightarrow$ ($\frac{h}{4}$,$\frac{w}{4}$,256) & ResBlock: CONV-($N$256, $K$3x3, $S$1, $P$1), AdaIN, ReLU \\ \hline\hline
\end{tabular}
\end{table*}

\begin{table*}[h]
\centering
\caption{Architecture of MLP for encoder $\gene$ and decoder $\gend$. One single style representation for each domain is learned from the domain label $d_i$.}
\begin{tabular}{ccc}
Part                 & Input $\rightarrow$ Output Shape & Layer Information \\\hline\hline
\multirow{3}{*}{MLP} & (2) $\rightarrow$ (256)                   & FC(2, 256), ReLU   \\
                     & (256) $\rightarrow$ (256)                 & FC(256, 256), ReLU \\
                     & (256) $\rightarrow$ ($n_s$)                  & FC(256, $n_s$)  \\ \hline\hline
\end{tabular}
\end{table*}

\begin{table*}[h]
\centering
\caption{Architecture of the unified decoder $\gend$. The style representation learned from MLP for each domain is injected by adaptive instance normalization.}
\begin{tabular}{ccc}
Part                           & Input $\rightarrow$ Output Shape  & Layer Information                  \\ \hline\hline
\multirow{4}{*}{Bottleneck}    & ($\frac{h}{4}$,$\frac{w}{4}$,256)+(256+256) $\rightarrow$ ($\frac{h}{4}$,$\frac{w}{4}$,256) & ResBlock: CONV-($N$256, $K$3x3, $S$1, $P$1), AdaIN, ReLU \\
                               & ($\frac{h}{4}$,$\frac{w}{4}$,256)+(256+256) $\rightarrow$ ($\frac{h}{4}$,$\frac{w}{4}$,256) & ResBlock: CONV-($N$256, $K$3x3, $S$1, $P$1), AdaIN, ReLU \\
                               & ($\frac{h}{4}$,$\frac{w}{4}$,256)+(256+256) $\rightarrow$ ($\frac{h}{4}$,$\frac{w}{4}$,256) & ResBlock: CONV-($N$256, $K$3x3, $S$1, $P$1), AdaIN, ReLU \\
                               & ($\frac{h}{4}$,$\frac{w}{4}$,256)+(256+256) $\rightarrow$ ($\frac{h}{4}$,$\frac{w}{4}$,256) & ResBlock: CONV-($N$256, $K$3x3, $S$1, $P$1), AdaIN, ReLU \\ \hline
\multirow{3}{*}{Up-sampling} & ($\frac{h}{4}$,$\frac{w}{4}$,256)+(128+128) $\rightarrow$ ($\frac{h}{2}$,$\frac{w}{2}$,128) & DECONV-($N$128, $K$3x3, $S$2, $P$1), AdaIN, ReLU            \\
                               & ($\frac{h}{2}$,$\frac{w}{2}$,128)+(64+64) $\rightarrow$ (h,w,64)                          & DECONV-($N$64, $K$3x3, $S$2, $P$1), AdaIN, ReLU           \\
                               & (h,w,64)$\rightarrow$(h,w,1)                                                    & CONV-($N$1,$K$7x7,$S$1,$P$3), ReLU, Tanh          \\ \hline\hline
\end{tabular}
\end{table*}

\begin{table*}[h]
\centering
\caption{Architecture of domain discriminator $\dis$.}
\begin{tabular}{ccc}
Part                         & Input $\rightarrow$ Output Shape & Layer Information \\ \hline\hline
\multirow{4}{*}{Shared body} & (h,w,1) $\rightarrow$ ($\frac{h}{2}$,$\frac{w}{2}$,64) & CONV-($N64$, $K$4x4, S2, P1), LReLU \\
                             & ($\frac{h}{2}$,$\frac{w}{2}$,64) $\rightarrow$ ($\frac{h}{4}$,$\frac{w}{4}$,128) & CONV-($N$128, $K$4x4, $S$2, $P$1), IN, LReLU \\
                             & ($\frac{h}{4}$,$\frac{w}{4}$,128) $\rightarrow$ ($\frac{h}{8}$,$\frac{w}{8}$,256)& CONV-($N$256, $K$4x4, $S$2, $P$1), IN, LReLU \\
                             & ($\frac{h}{8}$,$\frac{w}{8}$,256) $\rightarrow$ ($\frac{h}{8}$,$\frac{w}{8}$,512) & CONV-($N$512, $K$4x4, $S$1, $P$1), IN, LReLU \\ \hline
Output Branch $D_1$                     & ($\frac{h}{8}$,$\frac{w}{8}$,512) $\rightarrow$ ($\frac{h}{8}$,$\frac{w}{8}$,1) & CONV-($N$1, $K$4x4, $S$1, $P$1) \\ \hline
Output Branch $D_2$                    & ($\frac{h}{8}$,$\frac{w}{8}$,512) $\rightarrow$ ($\frac{h}{8}$,$\frac{w}{8}$,1) & CONV-($N$1, $K$4x4, $S$1, $P$1) \\ \hline\hline
\end{tabular}
\end{table*}

\newpage
\subsection{Adaptive Instance Normalization (AdaIN)}
Instance Normalization is defined in the following:
\begin{equation}
IN(x)=\gamma(\frac{x-\mu(x)}{\sigma(x)})+\beta
\end{equation}
where $\gamma$ and $\beta$ are learnable parameters, and $\mu(x)$ and $\sigma(x)$ are the channel-wise mean and variance which are defined as:
\begin{equation}
\mu_{nc}(x)=\frac{1}{HW}\sum_{h=1}^{H}\sum_{w=1}^{W} x_{nchw}, \sigma_{nc}(x)=\sqrt{\frac{1}{HW}\sum_{h=1}^{H}\sum_{w=1}^{W} (x_{nchw}-\mu_{nc}(x))}
\end{equation}

According to previous works~\cite{StarGANv2,MUNIT,adain}, AdaIN receives a content input $x$. And a style representation $y$ contains $y_\mu$ and $y_\sigma$. Unlike Instance Normalization (IN), AdaIN has no learnable affine parameters. Instead, it  uses adaptively the input style representations as the affine parameters:
\begin{equation}
AdaIN(x,y_\mu,y_\sigma)=y_\mu(\frac{x-\mu(x)}{\sigma(x)})+y_\sigma.
\end{equation}

More details can be seen in the codes.

\subsection{Implementation Details}
\subsubsection{Image Preprocessing}
All the datasets are scaled using min-max normalization to [-1, 1] directly. 3D data are then spatially normalized to have the same spacing on the three axis. 2D patches or 3D sub-volumes are randomly cropped during training. For some 3D datasets, we resize the normalized images with the ratio of 0.5 to reduce their computational redundancy, as shown in \autoref{tab:msyn}.

\subsubsection{Mask Synthesis}
To minimize the influence of mask synthesis on the evaluation of unsupervised methods, we utilize the simplest simulation model to describe the nucleus. More concretely, we use the ellipse and ellipsoid to represent nuclei in 2D and 3D, respectively. To generate such masks, three properties are typically required: ellipse/ellipsoid per image ($P_n$), the size of ellipse/ellipsoid ($P_s$), and the eccentricity of ellipse/ellipsoid ($P_e$). To imitate real unsupervised training, the range of nuclei number $n$ in each image is roughly estimated.  $P_s$ are tuned manually by determining the range of each ellipse/ellipsoid's major axis $a$. For $P_e$, the eccentricity $e$ is randomized in the range of [0.25, 0.75] for all the experiments to calculate the minor axis $b$ by the equation:
\begin{equation}
b=\sqrt{1-e^2}a
\end{equation}
The generated ellipse/ellipsoid are then randomly rotated and placed in the masks without overlapping. Details of parameters setting can be found in \autoref{tab:msyn}.

\begin{table}[!h]
\centering
\caption{Detailed parameters used to synthesize masks for all experiments.}
\begin{tabular}{lcccc}
Dataset               &Resize Ratio& Major axis of ellipse/ellipsoid $a$  &  Number of ellipse/ellipsoid $n$  &Synthetic Patch Size \\\hline\hline
Fluo-N2DL-HeLa (semantic segmentation) &1.0& {[}13, 18{]} & {[}5, 40{]}    &    [256,256]          \\
HaCaT                 &1.0& {[}20, 30{]} & {[}5, 15{]}    &    [256,256]          \\
BBBC024               &1.0& {[}30, 35{]} & {[}13, 15{]}   &     [128,512,768]        \\
Scaffold-A549        &0.5 & {[}10, 15{]} & {[}50, 100{]} &     [64,128,128]          \\
Fluo-N2DL-HeLa (instance segmentation)   &1.0& {[}13, 18{]} & {[}10, 40{]}   &     [256,256]      \\
Fluo-N2DH-SIM+        &1.0& {[}23, 35{]} & {[}5, 13{]}    &    [256,256]         \\
Fluo-N2DH-GOWT1      &1.0& {[}15, 23{]} & {[}4, 12{]}    &    [256,256]          \\
Fluo-N3DH-CE         &0.5 & {[}10, 18{]} & {[}2, 60{]}    &    [256,256]      \\
Fluo-N3DH-CHO        &1.0 & {[}15, 30{]} & {[}6, 16{]}    &    [64,128,128]          \\
Fluo-N3DH-SIM+       &0.5 & {[}10, 15{]} & {[}2, 24{]}    &    [64,128,128]          \\
 \hline\hline
\end{tabular}
\label{tab:msyn}
\end{table}
%Fluo-N3DL-TRIC       &0.5& {[}8, 27{]}  & {[}15, 30{]}   &    [64,128,128]         \\

\newpage
\subsection{Two-stage Pipeline with AD-GAN}
We conduct further comparison of the proposed AD-GAN-INS and a pipeline with a state-of-the-art supervised medical image method TransUNet~\cite{chen2021transunet} trained on synthetic AD-GAN training data.
As shown in \autoref{tab:pip}, we conduct experiments under different setting.
% First, we train the TransUNet with the synthetic paired data from CycleGAN which contains nuclei offset, nuclei deletion/addition, and inconsistent shape.

First, we train the TransUNet with the synthetic paired data from CycleGAN which contains nuclei offset, nuclei deletion/addition, and inconsistent shape. Compared with CGU-Net which utilizes U-Net as segmentor, an improvement of 5.5\% on DICE can be observed, showing the importance of using a robust model under erroneous data. However, compared with our method, there is still a large gap  mainly because data synthesized by CycleGAN are significantly of lower quality than those by our method.

Second, we also show that  training TransUNet with the synthetic paired data from AD-GAN can lead to a further improvement of 1.8\% and 0.9\% on segmentation and detection respectively, compared with baseline AD-GAN-INS. Using a two-stage pipeline with AD-GAN shall bring steady improvement but with extra computation cost and more training time, which can be seen as a trade-off between efficiency and accuracy.

Finally, a gap between the supervised TransUNet and the two-stage pipeline can be noticed. This might be partly caused by the simple simulation of nuclei in our pipeline, which may not reflect the diversity and complexity of nuclei shape. Nonetheless, considering the unsupervised nature of our proposed method which need no annotation, $91.1\%$ or $89.3\%$ can still be deemed pretty good.  Future work may explore the possibility of generating data more realistically to minimize the gap between supervised methods and unsupervised methods.

% Compared with CGU-Net which utilizes U-Net as the segmentor, an improvement of 5.5\% on DICE can be observed; this shows  the importance of using a robust model under erroneous data. However, it is still much lower than our method due to caused by the misleading of incorrect data. After that, training TransUNet with the synthetic paired data from AD-GAN led to an improvement of 1.8\% and 0.9\% on segmentation and detection respectively, compared with baseline AD-GAN-INS. Using a two-stage pipeline with AD-GAN shall bring up steady improvement along with extra computation cost and training time, which can be seen as a trade-off between efficiency and accuracy. Meanwhile, a large gap between the supervised TransUNet and the two-stage pipeline can be noticed. It may be caused by the too simple simulation of nuclei in our pipeline, which cannot reflect the diversity and complexity of nuclei shape.

\begin{table}[!h]
\centering
\caption{Comparison of AD-GAN-INS and two-stage AD-GAN pipeline on dataset HaCaT.}
\begin{tabular}{|l|l|c|c|}
\hline
Model & Training & DICE (segmentation) & F1-score (detection) \\ \hline
AD-GAN-INS & unsupervised & 89.3 & 95.2 \\ \hline
TransUNet & CycleGAN synthetic data & 78.8 & 88.4 \\ \hline
TransUNet & AD-GAN synthetic data & 91.1 & 96.1 \\ \hline
TransUNet & Annotated data & 95.2 & 97.5 \\ \hline
\end{tabular}
\label{tab:pip}
\end{table}

\subsection{Interpolation of Domain Label}
\begin{figure}[h]
\begin{center}
%\fbox{\rule{0pt}{1in} \rule{0.9\linewidth}{0pt}}
\includegraphics[width=1\linewidth]{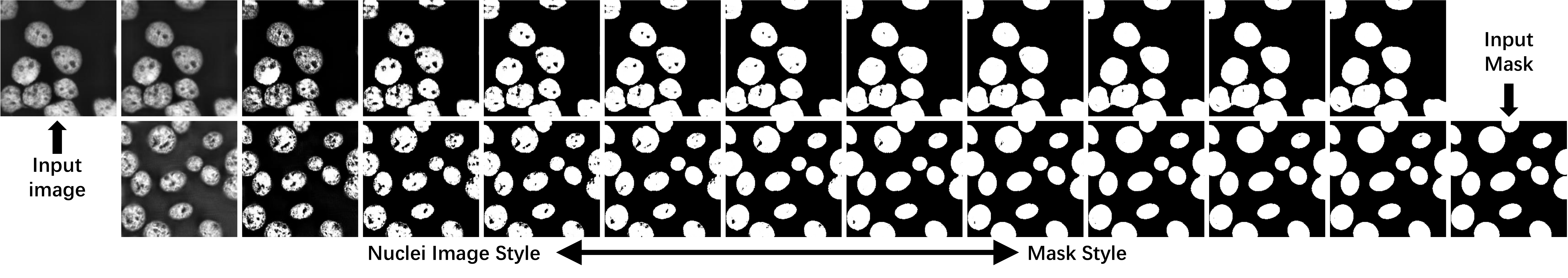}
\end{center}
\caption{Continuous change of style representation}
\label{fig:cc}
\end{figure}
According to UFDN~\cite{UFDN}, if a model can really learn the representation disentanglement, the generated images should show a continuous change when the input style representation is changed accordingly. In order to validate this, we generate continuous change of style representation by linearly interpolating the two domain labels between (0,1) and (1,0) with 10 intervals. We then input them to the trained AD-GAN (\autoref{fig:cc}).

\newpage
\subsection{Cell Tracking Challenge}
\begin{table}[!h]
\caption{Comparison of AD-GAN-INS and the state-of-the-art methods on Cell Segmentation Benchmark. CycleGAN fails to establish one-to-one mapping in some datasets and hence we leave their results as ``-".}
\begin{tabular}{cccccccc}
\begin{tabular}[c]{@{}c@{}}\end{tabular} & Method   & Fluo-N2DH-GOWT1 & Fluo-N2DL-HeLa & Fluo-N2DH-SIM+ & Fluo-N3DH-CE & Fluo-N3DH-CHO & Fluo-N3DH-SIM+ \\\hline
\multirow{5}{*}{$\rm{OP}_{\rm{csb}}  $ }                                          & First    & 0.952           & 0.954          & 0.905          & 0.83         & 0.926         & 0.949          \\
                                                                & Second   & 0.948           & 0.953          & 0.897          & 0.816        & 0.913         & 0.885          \\
                                                                & Third    & 0.948           & 0.951          & 0.897          & 0.811        & 0.913         & 0.866          \\
                                                                & CycleGAN & -             & 0.654          & 0.632          & -          & 0.644         & -            \\
                                                                & AD-GAN   & 0.804           & 0.894          & 0.824          & 0.729        & 0.852         & 0.733          \\\hline
\multirow{5}{*}{SEG}                                            & First    & 0.938           & 0.923          & 0.832          & 0.729        & 0.917         & 0.906          \\
                                                                & Second   & 0.933           & 0.919          & 0.825          & 0.705        & 0.914         & 0.786          \\
                                                                & Third    & 0.931           & 0.917          & 0.822          & 0.688        & 0.903         & 0.759          \\
                                                                & CycleGAN & -             & 0.486          & 0.457          & -          & 0.506         & -            \\
                                                                & AD-GAN   & 0.787           & 0.850          & 0.720          & 0.624        & 0.823         & 0.613          \\\hline
\multirow{5}{*}{DET}                                            & First    & 0.98            & 0.994          & 0.983          & 0.917        & 0.954         & 0.992          \\
                                                                & Second   & 0.976           & 0.992          & 0.981          & 0.914        & 0.945         & 0.984          \\
                                                                & Third    & 0.970           & 0.992          & 0.979          & 0.903        & 0.934         & 0.974          \\
                                                                & CycleGAN & -             & 0.823          & 0.806          & -          & 0.782         & -            \\
                                                                & AD-GAN   & 0.821           & 0.938          & 0.926          & 0.833        & 0.881         & 0.852          \\\hline
\end{tabular}
\label{tab:allctc}
\end{table}

We conduct extra experiments of instance segmentation on Cell Tracking Challenge, and directly compare our method with the state-of-the-art methods on Cell Segmentation Benchmark\footnote{The benchmark can be found in \href{http://celltrackingchallenge.net/files/leaderboards/CSB/2021-10-22.png}{Cell Segmentation Benchmark}}. Particularly, we evaluate our method on Fluo-N2DH-GOWT1,  Fluo-N2DL-HeLa, Fluo-N2DH-SIM+, Fluo-N3DH-CE, Fluo-N3DH-CHO, and Fluo-N3DH-SIM+. We follow the official metrics to report the performance by SEG, DET, and $\rm{OP}_{\rm{csb}}$, where $\rm{OP}_{\rm{csb}}= 0.5\times(\rm{SEG}+\rm{DET})$.
We also report the results from CycleGAN, as a baseline of unsupervised nuclei segmentation. The training protocol is the same for AD-GAN-INS and CycleGAN, including mask synthesis and maximum training iteration.

Overall, CycleGAN fails to learn one-to-one mapping in the datasets Fluo-N2DH-GOWT1, Fluo-N3DH-CE, and Fluo-N3DH-SIM+. That is to say, the trained CycleGAN cannot generate masks depending on the input image, but generates masks randomly, which may be caused by the difficulty of building one-to-one mapping as well as lacking  constraints in the training process of CycleGAN. In addition, due to the lossy transformation problem, CycleGAN achieves low performance in segmentation tasks (caused by nuclei offset, inconsistent shape) on the other three datasets. Meanwhile, the problem of nuclei addition/deletion also has an influence on detection results of CycleGAN, resulting in a low score of $\rm{OP}_{\rm{csb}}$.

In comparison, AD-GAN-INS achieves stable and reasonable results on all the six datasets.  Nonetheless, it should be admitted that there is still a gap ($0.113\pm0.054$) between our unsupervised method and the top ranked  methods in the lead board of CTC.  On one hand, we believe this is understandable, as the proposed AD-GAN-INS is unsupervised: without using any annotated data, the performance of around 80\% (even on some hard datasets) appears already quite promising. By contrast, most of the top ranked methods in CTC are supervised methods (pretrained on ImageNet, and requiring heavy annotation). Though a few leading methods are image-processing based approaches, they rely on task-specific complicated image processing techniques (including sophisticated pre- and post-processing). They are hard to be adapted to the other tasks and difficult to be used by practitioners.  On the other hand,  we also argue that it may not be very fair to evaluate our method on CTC, which contains only two annotated videos in each dataset.  As unsupervised methods may need more training samples to catch up with supervised methods, insufficient training samples actually limit our AD-GAN's performance. We believe it may be more suitable for our methods to be applied on larger un-annotated datasets where AD-GAN can become much powerful.

\newpage
\subsection{Scaffold-A549 dataset for Unsupervised Nuclei Segmentation}
\begin{figure*}[htb]
\centering
\includegraphics[width=0.95\columnwidth]{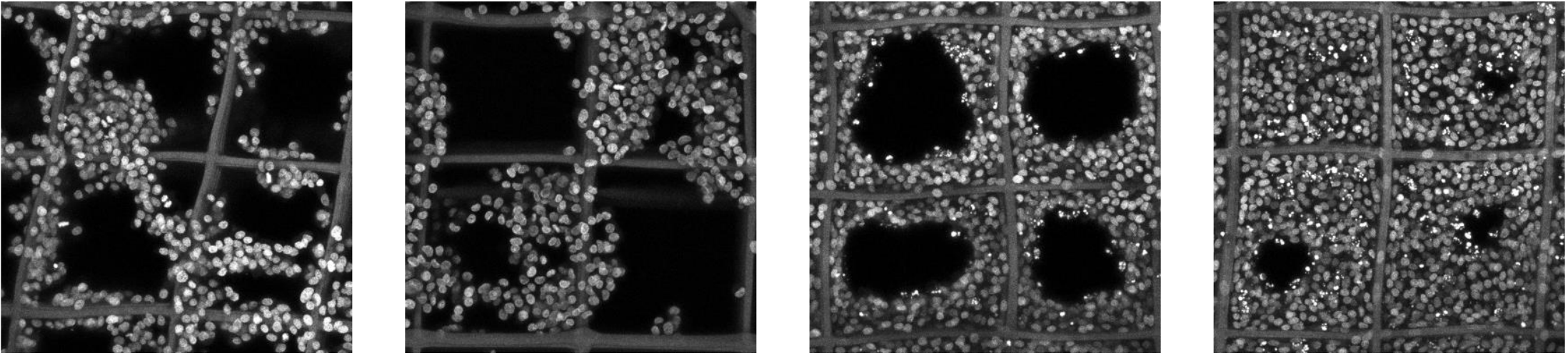}
\caption{Visualization of samples from the scaffold-A549 datasets.}
\label{fig:a549}
\end{figure*}

 As part of this work, we introduce a new 3D fluorescence image dataset that we term as scaffold-based A549 cell culture (Scaffold-A549) dataset\footnote{Scaffold-A549 is available at: \href{https://github.com/Kaiseem/Scaffold-A549}{https://github.com/Kaiseem/Scaffold-A549}.}, which is more challenging. Scaffold-A549 consists of 20 unlabelled training images and one fully annotated test image. The A549 human non-small cell lung cancer cells were seeded in fibrous scaffolds and the nuclei of A549 cells are stained with Hoechst 33342 (blue) for Confocal laser scanning microscope (CLSM) imaging. Images of 3D cell cultures were captured with a EC Plan-Neofluar 20X/0.5 air immersion objective, using a CLSM system (LSM-880, ZEISS, Germany), and the scanning depth of the cell cultured scaffolds is set about 60$\mu m$ based on preliminary tests. Then, a spatial normalization and center crop operation is performed to all 21 collected CLSM with size of $1024\times1024\times128$ voxels and resolution of 0.4151 $\mu m$ per voxel. The density of nuclei varies among the images, and one image with the median density is adopted for manual annotation. To reduce the annotation ambiguity, this volume is labelled slice-by-slice from axial, sagittal and coronal views by three professional persons using 2D labelling software. A total of 9 annotations are merged using the average value, followed by a Gaussian filter for surface smoothing. Annotating the data in this way can ensure as accurate as possible the nuclear boundary in the annotation process, though the nucleus is elongated axially due to light diffraction. %However, it should be noted that pixel-level difference between annotation and the true nuclear boundary cannot be fully avoided in some challenging cases, e.g., the nuclei near the opaque scaffold. For our experiments, the cropped sub-volumes of unlabelled images are used to train the unsupervised model, and the annotation is used to evaluate the performance of each method. We display some example images from the Scaffold-A549 dataset in \autoref{fig:a549}.

\newpage

\subsection{Visualization of BBBC024 Results}
\begin{figure*}[ht]
\centering
\includegraphics[width=0.7\columnwidth]{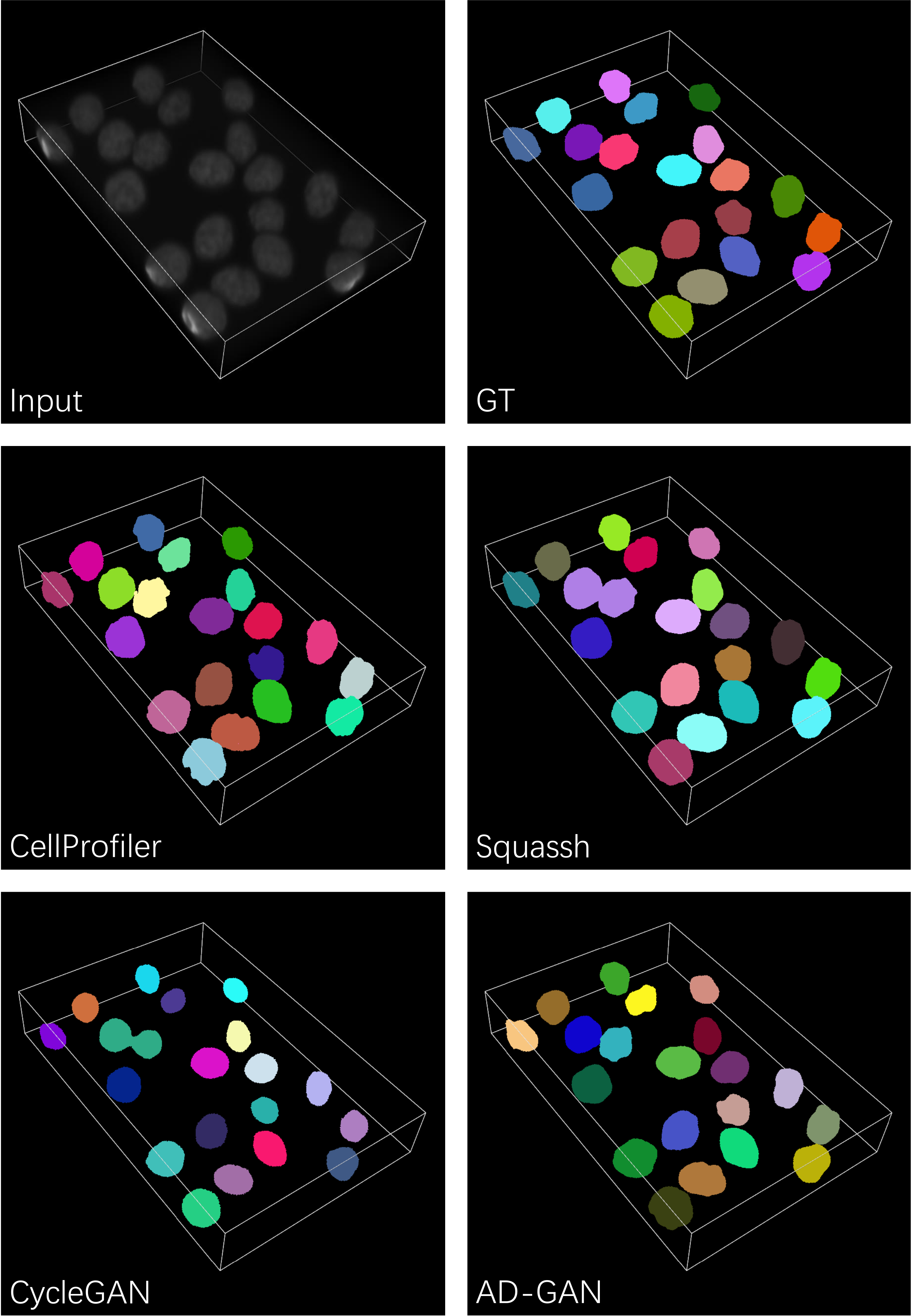}
\caption{Visualization of segmentation results on 3D data BBBC024.}
\end{figure*}

\newpage

\subsection{Parameter Selection in Synthesizing Masks }
\begin{figure*}[ht]
\centering
\includegraphics[width=0.95\columnwidth]{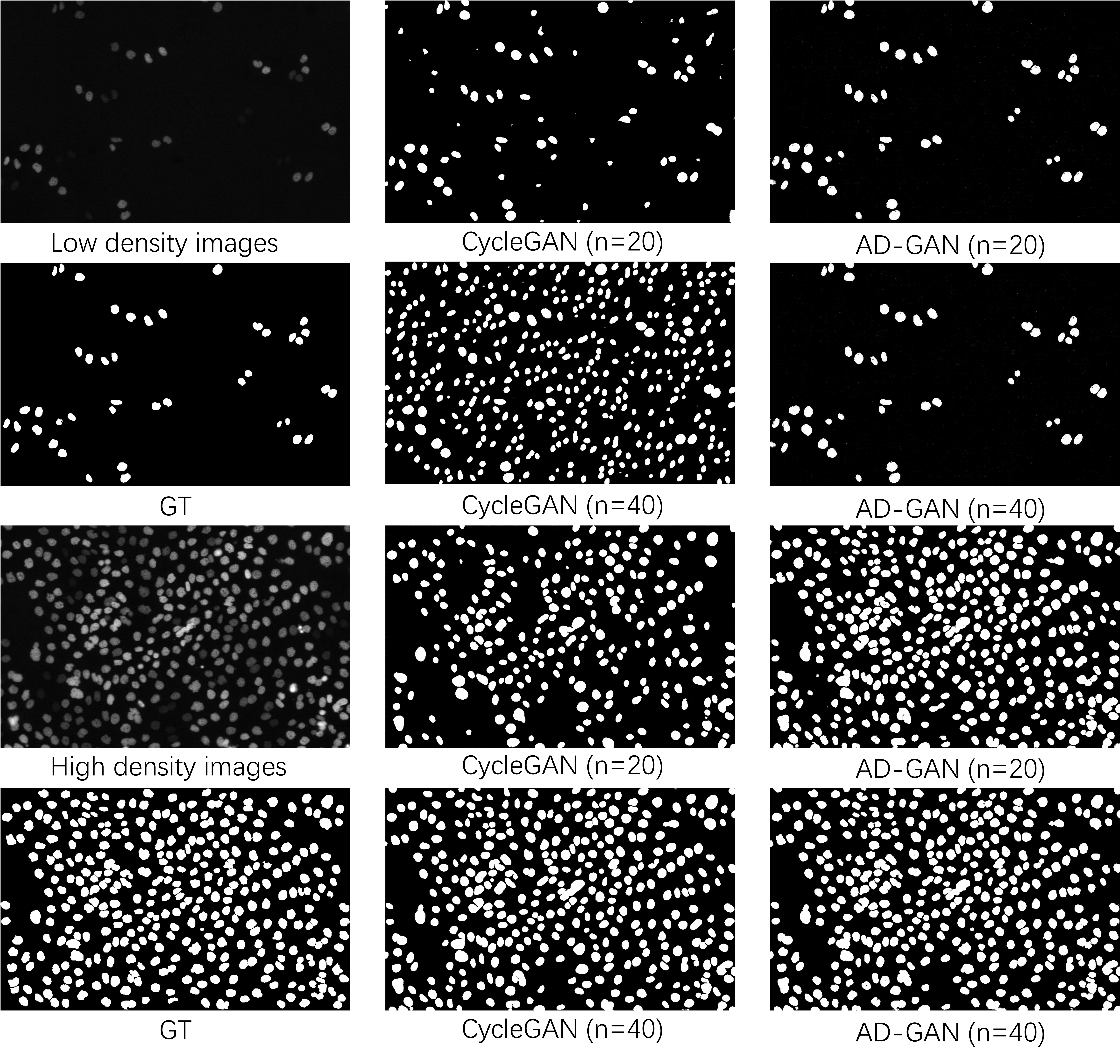}
\caption{Comparison results on image-to-mask translation with different  nuclei number $n$ in a synthetic mask.}
\label{fig:nc2}
\end{figure*}

\newpage

\subsection{Visualization of Generative Results }
\begin{figure*}[ht]
\centering
\includegraphics[width=0.95\columnwidth]{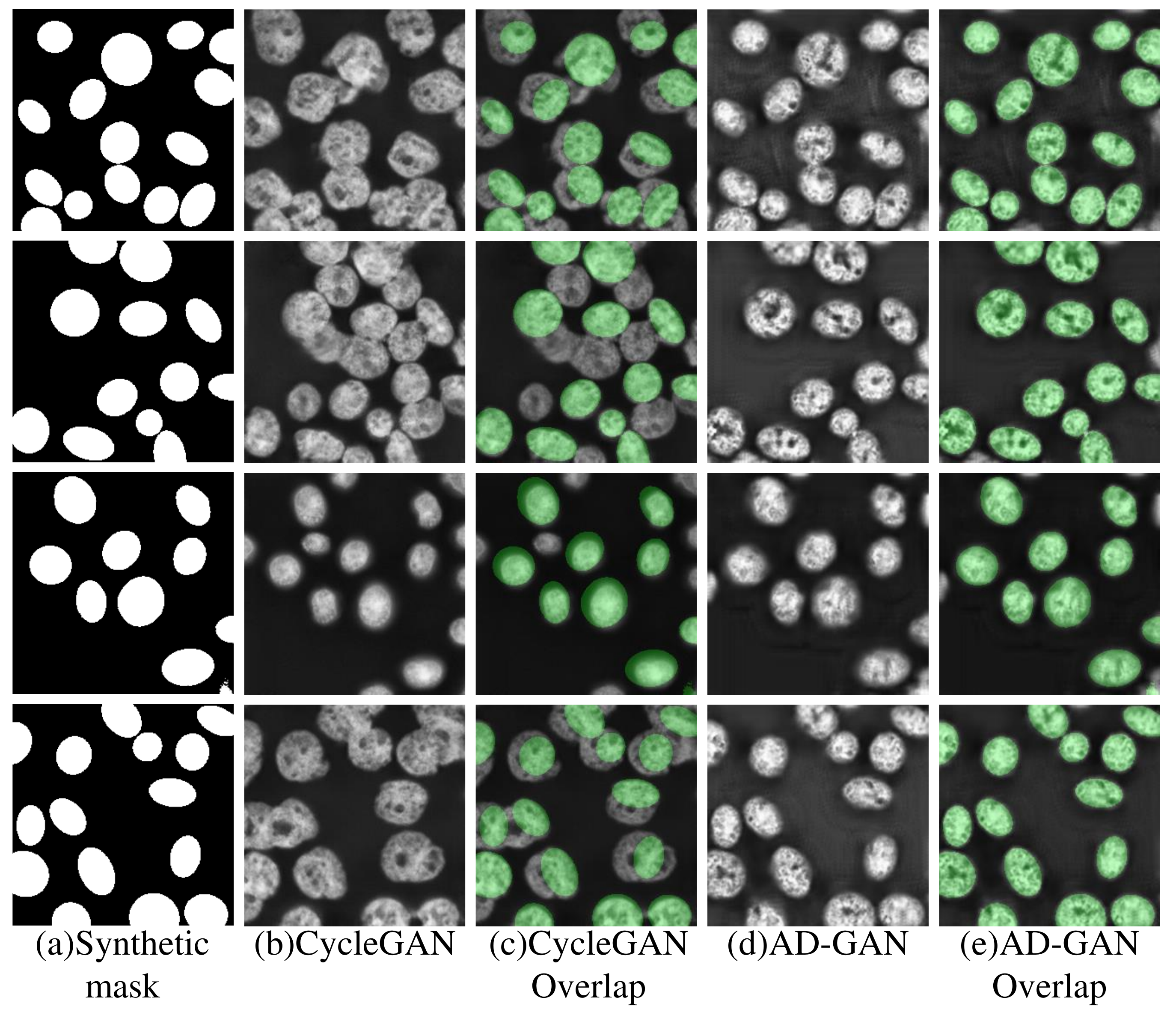}
\caption{Generative results comparison from mask to images on datasets HaCaT. (a): The randomly synthetic masks; (b) and (d): CycleGAN's and AD-GAN's generated images from (a); (c) and (e):  overlapped images containing both synthetic mask and generated images. It is argued here that two-stage pipelines cannot guarantee good segmentation performance if the semantic object one-to-one mapping cannot be learned well at the first stage. More online generation results can be shown in the codes by changing the seed.}
\end{figure*}

\end{document}